\pdfoutput=1

	\def\clsstyle{prb}

	\newcommand{\prltext}[1]{}

    \newcommand{\ket}[1]{| #1 \rangle}

    \newcommand{\Tr}{\text{Tr}}

\documentclass[aps,\clsstyle,reprint,twocolumn,showpacs,superscriptaddress,longbibliography]{revtex4-1}

\usepackage{amsmath, amssymb, bm, braket, dsfont, times, mathtools,graphicx,color}
\usepackage{natbib}
\usepackage[breaklinks,colorlinks,allcolors=blue]{hyperref}
\usepackage{tikz}
\usepackage[normalem]{ulem}
\usepackage{enumerate}
\usepackage{float}
\usepackage{multirow}
\usepackage{booktabs}

\AtBeginDocument{%
  \heavyrulewidth=.08em
  \lightrulewidth=.05em
  \cmidrulewidth=.06em
  \belowrulesep=.65ex
  \belowbottomsep=0pt
  \aboverulesep=.4ex
  \abovetopsep=0pt
  \cmidrulesep=\doublerulesep
  \cmidrulekern=.5em
  \defaultaddspace=.5em
}

\begin{document}

\title{Bootstrapping Lieb-Schultz-Mattis anomalies}
\author{Ryan A. Lanzetta}
\author{Lukasz Fidkowski}
\affiliation{Department of Physics, University of Washington Seattle}
\date{\today}
\begin{abstract}
We incorporate the microscopic assumptions that lead to a certain generalization of the Lieb-Schultz-Mattis (LSM) theorem for one-dimensional spin chains into the conformal bootstrap. Our approach accounts for the ``LSM anomaly'' possessed by these spin chains through a combination of modular bootstrap and correlator bootstrap of symmetry defect operators. We thus obtain universal bounds on the local operator content of (1+1)$d$ conformal field theories (CFTs) that could describe translationally invariant lattice Hamiltonians with a $\mathbb Z_N\times \mathbb Z_N$ symmetry realized projectively at each site. We present bounds on local operators both with and without refinement by their global symmetry representations. Interestingly, we can obtain non-trivial bounds on charged operators when $N$ is odd, which turns out to be impossible with modular bootstrap alone. Our bounds exhibit distinctive kinks, some of which are approximately saturated by known theories and others that are unexplained. We discuss additional scenarios with the properties necessary for our bounds to apply, including certain multicritical points between (1+1)$d$ symmetry protected topological phases, where we argue that the anomaly studied in our bootstrap calculations should emerge.

\end{abstract}

\maketitle
\tableofcontents

\section{Introduction}

\subsection{Overview}
Identifying the low energy spectrum of a given lattice Hamiltonian is an important goal of quantum many-body theory. In certain cases, depending on the symmetries of the model, the qualitative nature of its spectrum can be constrained, thus restricting the potential quantum field theory (QFT) descriptions. A famous and powerful result of this kind, known as the Lieb-Schultz-Mattis (LSM) theorem, is that half-odd-integer spin Heisenberg chains are gapless\cite{Lieb:1961fr}. This is to be contrasted with the Haldane gap for the case of integer spin\cite{Haldane:1983ru,Affleck:1986pq}. Following these results, various generalizations in similar spirit have been made: in higher dimensions\cite{cond-mat/9911137,Hastings:2003zx}, with more generic spatial symmetries \cite{1212.0557,1505.04193,PhysRevB.96.205106,Else:2019lft}, including the magnetic translation group \cite{Cheng:2018lti, Lu:2017ego, Yang:2017frp}, and with higher form symmetries\cite{Kobayashi:2018yuk}. In this work, we will be concerned with an extension of the LSM theorem to general global symmetry groups, where it is expected that a translationally invariant local spin chain with an on-site global symmetry represented projectively at each site cannot be trivially gapped\cite{PhysRevB.83.035107,math-ph/1808.08740,Ogata:2020hry,Prakash:2020hje}.  This leaves gaplessness or spontaneous symmetry breaking (SSB) as the only possibilities in one spatial dimension. \par 

The modern formulation of LSM-type theorems is in terms of 't Hooft anomalies\cite{PhysRevLett.118.021601,Cheng:2015kce,PhysRevB.96.195105, PhysRevB.98.085140, Cordova:2019bsd}, which can be viewed as obstructions to gauging a global symmetry. For a lattice model subject to the generalized LSM theorem with an internal symmetry $G$, what we will refer to as the \emph{LSM anomaly} is a mixed anomaly between lattice translation symmetry and $G$. This anomaly arises due to the fact that inserting a defect of translation symmetry amounts to adding one site, which carries a projective representation of $G$.  Thus, in the presence of such translation symmetry defects, it is not possible to gauge $G$.  This 't Hooft anomaly must be matched by both the lattice and QFT descriptions\cite{tHooft:1979rat}, making it a powerful non-perturbative tool to aid in identifying candidate low energy theories––typically a complicated task.\par 

Within the realm of QFT, another set of non-perturbative techniques are those related to conformal field theory (CFT), among them being the numerical conformal bootstrap. Following recent work combining $\mathbb Z_N$ symmetries and anomalies with modular bootstrap\cite{Lin:2019kpn,Lin:2021udi}, in this work we will use conformal bootstrap to bound the space of CFTs that possess certain LSM anomalies arising in lattice models with global, internal symmetry $G_{\text{int}} = \mathbb Z_N^2$. We will assume, additionally, that the lattice translation symmetry is realized as a $\mathbb Z_N$ internal symmetry in the CFT, leading the CFTs we consider to have a minimal, internal symmetry group $G = \mathbb Z_N^3$. Incorporating the signatures of the LSM anomalies into bootstrap is somewhat subtle; to do so, we introduce a new technique that augments modular bootstrap by incorporating additional numerical bounds that come from imposing crossing symmetry on four-point functions of certain symmetry defect operators. This approach allows us to obtain universal bounds on the local operator content of 1+1$d$ CFTs in a way that is refined by LSM anomalies.\par

\subsection{Background, Methods and Motivation}

Lattice models with the kinds of symmetries and anomalies we have mentioned have received some recent attention, partially motivating this work. Under the assumption of a unique ground state, CFTs naturally describe gapless spin chains satisfying LSM constraints, since any scale-invariant, (1+1)$d$ QFT is necessarily a CFT, under mild assumptions\cite{Zamolodchikov:1986gt,Polchinski:1987dy,Cardy:1996xt,Nakayama:2013is}. Indeed, in some recent numerical simulation work it was observed that entire stable, gapless phases of translation-invariant spin chains, subject to LSM constraints with on-site $\mathbb Z_N \times \mathbb Z_N$ symmetries, are effectively described by theories within the conformal manifolds of $N-1$ compact bosons for $N=2,3$\cite{Alavirad:2019iea}. It was argued in  Ref. [\onlinecite{Alavirad:2019iea}] that a similar compact boson description should be valid for arbitrary, odd $N$. Further, the authors of  Ref. [\onlinecite{Alavirad:2019iea}] suggest that the central charge $c = N-1$ of the compact boson theories may be the minimum necessary to accommodate the LSM anomaly when the only microscopically-imposed symmetry is $\mathbb Z_N^3$, assuming that the $\mathbb Z_{\text{trans}}$ is realized as $\mathbb Z_N$ in the low energy global symmetry group $G_{\text{IR}}$. As noted by Ref. [\onlinecite{Cheng:2020rpl}], which also contains some discussion of the LSM anomalies studied in this work, if one imposes instead a larger $PSU(N)$ internal symmetry,  which is apparently emergent in a subset of the models studied by  Ref. [\onlinecite{Alavirad:2019iea}], then this minimum central charge is indeed $c=N-1$ by the Sugawara construction\cite{DiFrancesco}. However, as we will point out, there are trivial counterexamples to this bound when the minimum symmetry imposed at low energy is $\mathbb Z_N^3$  and $N$ is a product of coprime integers. Nonetheless, with a suitable quantitative modification to the possible central charge bound in these cases, there persists the difficult problem of determining whether non-trivial counterexamples exist. Thus, one of the goals of this work will be to look for bootstrap signatures of potential theories with a lower central charge that could have the LSM anomalies. \par
One possible explanation for the lack of counterexamples to the suggested central charge bound is that the space of (1+1)$d$ CFTs remains largely uncharted territory, with the vast majority of explicit constructions either being rational CFTs\cite{Moore:1988qv} (RCFTs) with enhanced symmetry or the minimal models\cite{Belavin:1984vu,PhysRevLett.52.1575}. Resurrecting ideas used originally to exactly solve many of the known examples of (1+1)$d$ CFTs, in the past several years it was realized that instead of attempting full solutions of specific CFTs, a still powerful and more tractable goal is to attempt to rule out, using numerical optimization techniques such as linear or semidefinite programming, certain regions in the space of all possible CFTs\cite{Rattazzi:2008pe,El-Showk:2012cjh,Poland:2018epd}—this represents the modern, numerical conformal bootstrap program. This is possible due to unitarity and other more stringent mathematical consistency requirements inherent to CFTs. In general dimensions, the main consistency requirement is the crossing symmetry of four-point functions. Within numerical bootstrap, one can attempt to show that certain assumptions about the spectrum of a CFT can lead to incompatibility with crossing symmetry. This approach, termed {\emph{correlator bootstrap}}, has had much success, and in some cases has gone so far as to produce numerical solutions to certain theories, as has been done in the case of the 3$d$ Ising CFT\cite{El-Showk:2012cjh,El-Showk:2014dwa,Simmons-Duffin:2016wlq}. These advances have especially been made possible following the introduction of specialized semidefinite programming packages for conformal bootstrap applications such as \texttt{SDPB}\cite{Simmons-Duffin:2015qma}. \par
In the setting of (1+1)$d$ CFTs, the possible constraints on the CFT data are more powerful than in $d > 2$ due to the correspondence between the local primary operator spectrum and the decomposition of the torus partition function into Virasoro characters, which is subject to modular invariance. Both analytically and numerically, it was shown that imposing modular invariance of the partition function gives generic bounds on the operator content of a unitary, compact, bosonic  (1+1)$d$ CFT, guaranteeing, for instance, an upper bound on the scaling dimension of the lightest primary field for any CFT\cite{Hellerman:2009bu,Collier:2016cls}. This should be contrasted with correlator bootstrap, where to obtain any bounds one must typically make an additional assumption that the theory possesses some fields with a particular scaling dimension\footnote{Actually, in general dimensions $d+1 > 2$ there are ways to put universal bounds on the local operator spectrum. Most generally, every unitary CFT possesses a conserved stress tensor, so it may be used as an external field in correlator bootstrap calculations and the content of its OPE with itself may be studied, leading to universal bounds on the lightest operator\cite{Dymarsky:2017yzx}. However, it is not possible to obtain universal bounds in this way that are refined by discrete, internal symmetries since the stress tensor does not interact non-trivially with such symmetries. For continuous internal symmetries, a similar approach may be used to obtain universal bounds by using the conserved currents of the symmetry\cite{Dymarsky:2017xzb}}. This approach, termed {\emph{modular bootstrap}}, has also been generalized to bosonic, (1+1)$d$ CFTs with anomalous and non-anomalous $\mathbb Z_N$ global symmetries by imposing modular covariance of the torus partition function twisted by symmetry defects \cite{Lin:2019kpn,Lin:2021udi}, as we mentioned. Additional progress in a similar vein has been made for fermionic CFTs with non-anomalous and anomalous global symmetries\cite{Benjamin:2020zbs,Grigoletto:2021zyv,Grigoletto:2021oho}, but in this work we will focus on bosonic theories. \par
The $\mathbb Z_N$ anomalies studied in previous modular bootstrap works are characterized by anomalous spin selection rules for so-called defect operators hosted at the end of topological defect lines (TDLs) implementing the $\mathbb Z_N$ symmetry\cite{Chang:2018iay}. These spin constraints lead to stronger unitarity bounds on the scaling dimensions for such operators. From these inputs emerges the general result that anomalous $\mathbb Z_N$ symmetries are necessarily accompanied by charged degrees of freedom at low energy. Further, the bounds on $\mathbb Z_N$-symmetric operators depend strongly on the anomaly. However, more complicated symmetry groups may have more complicated anomalies with more subtle signatures, ones that even do not include anomalous defect spin-selection rules.  Crucially for this work, a $\mathbb Z_N^3$ symmetry with LSM anomaly is one such case. As already alluded to, the main signature of the LSM anomaly is that symmetry defect operators of one $\mathbb Z_N$ subgroup transform in a projective representation of the remaining $\mathbb Z_N^2$ subgroup. Crucially, when $N$ is odd, this is essentially the only signature of the LSM anomaly; in these cases, there are no non-trivial defect spin selection rules, so modular bootstrap by itself is insensitive to the LSM anomaly and, thus, cannot give a bound on charged operators.  On the other hand, our approach gives rather tight bounds on charged operators at low central charge and uncovers various intriguing kinks. \par
To incorporate the LSM anomalies into bootstrap, we augment modular bootstrap by incorporating certain bounds coming from correlator bootstrap. 
Our approach exploits the constraining power of both crossing symmetry of defect operators and modular covariance of the twisted partition function as follows. Suppose we are trying to rule out some gap in the spectrum of scaling dimensions of local operators for CFTs with a particular central charge. In the case of the $\mathbb Z_N$ modular bootstrap, the presence of an anomaly sets \textit{universal} lower bounds on the scaling dimension of any $\mathbb Z_N$ symmetry defect operator; for the LSM anomaly, we derive a non-universal lower bound that depends on the assumed gap in the local operator spectrum and, in some cases, the central charge. The reason for this lower bound is that taking the operator product expansion (OPE) of light defect operators can produce light local operators; precisely how light the defect operators can be without necessarily producing a local operator whose scaling dimension violates the assumed gap in the local operator spectrum is quantified using correlator bootstrap. Additionally, in some cases the gaps in the spectrum of local and defect operators further lead to a lower bound on the central charge. This provides yet another route to improve our lower bound on the scaling dimension of the lightest defect operator, since the lower bound on the central charge must not be higher than the assumed central charge. On the other hand, for modular covariance of the twisted partition function to be obeyed, the lightest local operator and the lightest defect operator typically cannot both be too heavy; in particular, if the gap in the spectrum of local operators is large––for instance, a gap that we are trying to rule out––there often exists an upper bound on the gap in the spectrum of defect operators. This reasoning applies even when the gap among local operators is assumed to exist only in the charged sector, or only in the neutral sector. Modular bootstrap thus has the potential to rule out the combined gaps in the local and defect operator spectra, where the latter gap is implied by the former, leading to a contradiction and allowing us to rule out the assumed gap in the local operator spectrum. \par 
Our main results, shown in Figures \ref{fig:symmetric_plots} and \ref{main_plots}, are upper bounds, as a function of central charge $c$, on the lightest local operators with various symmetry properties for CFTs saturating the $\mathbb Z_N^3$ LSM anomaly for $N=2,3,4,5,6$. These bounds can be thought of as a refinement of the more qualitative LSM-type theorems, which only exclude a non-degenerate gapped ground state. Our results, by contrast, state that, for CFTs saturating the LSM anomalies, not only must there exist charged states with energies $\mathcal O(1/L)$ (in a ring geometry with periodic boundary conditions with $L$ the circumference and under the assumption of a unique ground state), but there is also a precise upper bound $\sim\Delta(c)/L$ on the energy of such states when the lattice model is described at low energy by a CFT with central charge $c$. There are various additional microscopic realizations of lattice models that can be described by the kinds of CFTs we put bounds on. These include certain multicritcal points of (1+1)$d$ symmetry protected topological (SPT) phases and edge theories of certain (2+1)$d$ SPT phases. \par 
\subsection{Organization}
The structure of the remainder of this paper is as follows. In section II we will present our universal bootstrap bounds on the local operator spectrum of (1+1)$d$ CFTs with the $\mathbb Z_N^3$ LSM anomalies for various $N$, and further discuss the implications of our bounds to the theory of multicritical points of SPT phase transitions. In section III we will provide technical background regarding symmetries and anomalies in (1+1)$d$ CFT, including details about the TDL formalism and how the LSM anomalies manifest within it. Then in section IV we will explain aspects of our numerical bootstrap approach and additionally present some of the other numerical bounds (Figures \ref{fig:scaling_bounds} and \ref{fig:cent_bounds}) that went into our final calculations. Finally, in section V we will make closing remarks and discuss potential future avenues for research. We provide additionally an appendix with details about our modular bootstrap calculations. \par  

\section{Main Results}
\begin{figure}[h]
    \centering
    \includegraphics[width=0.45\textwidth]{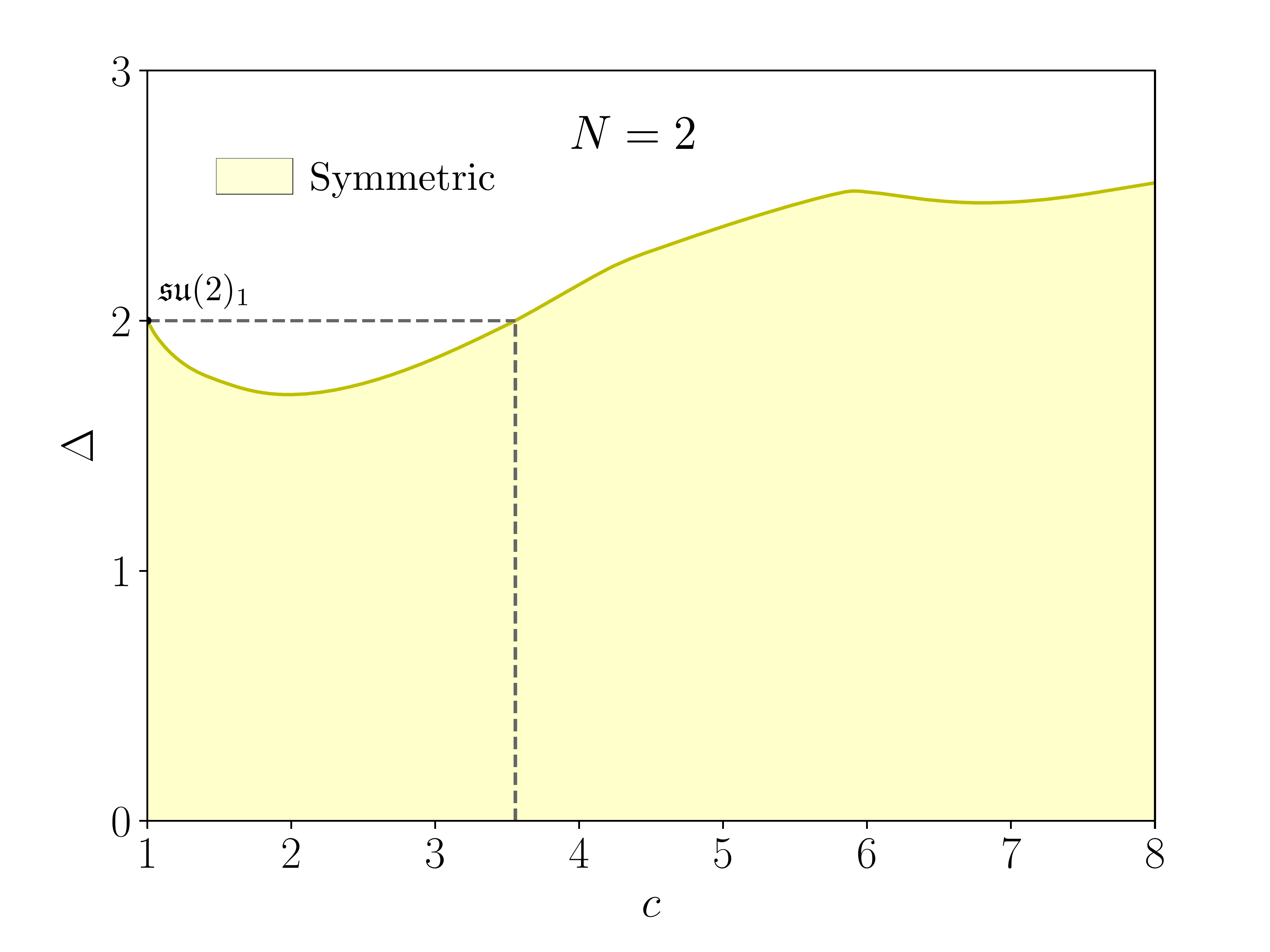}
    \includegraphics[width=0.45\textwidth]{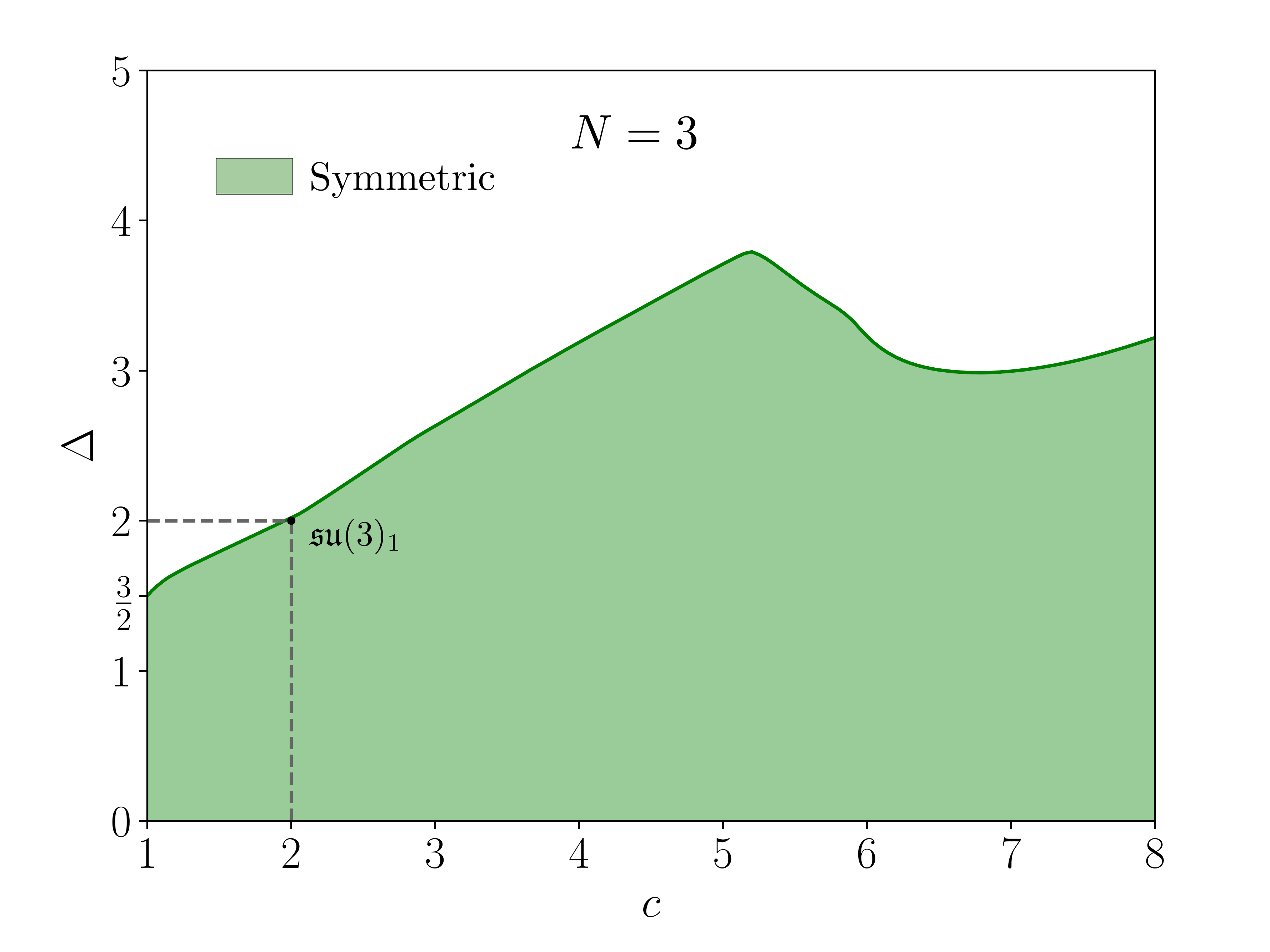}
    \caption{Upper bounds on the scaling dimension of the lightest $\mathbb Z_N^3$-symmetric scalar operator in a theory with the LSM anomaly, as a function of central charge, for $N=2,3$. The shaded region in each plot is thus the allowed region for the lightest symmetric, scalar operator. In obtaining these bounds, we did not make use of the additional improvements to standard modular bootstrap with global symmetry that we introduce in this work. The bounds were computed with $\Lambda^{\text{mod}} = 25$ and $S^{\text{mod}}_{\text{max}} = 50$ (see Section IV for implementation details).}
    \label{fig:symmetric_plots}
\end{figure}
\begin{figure*}
\centering
\includegraphics[width=0.458\textwidth]{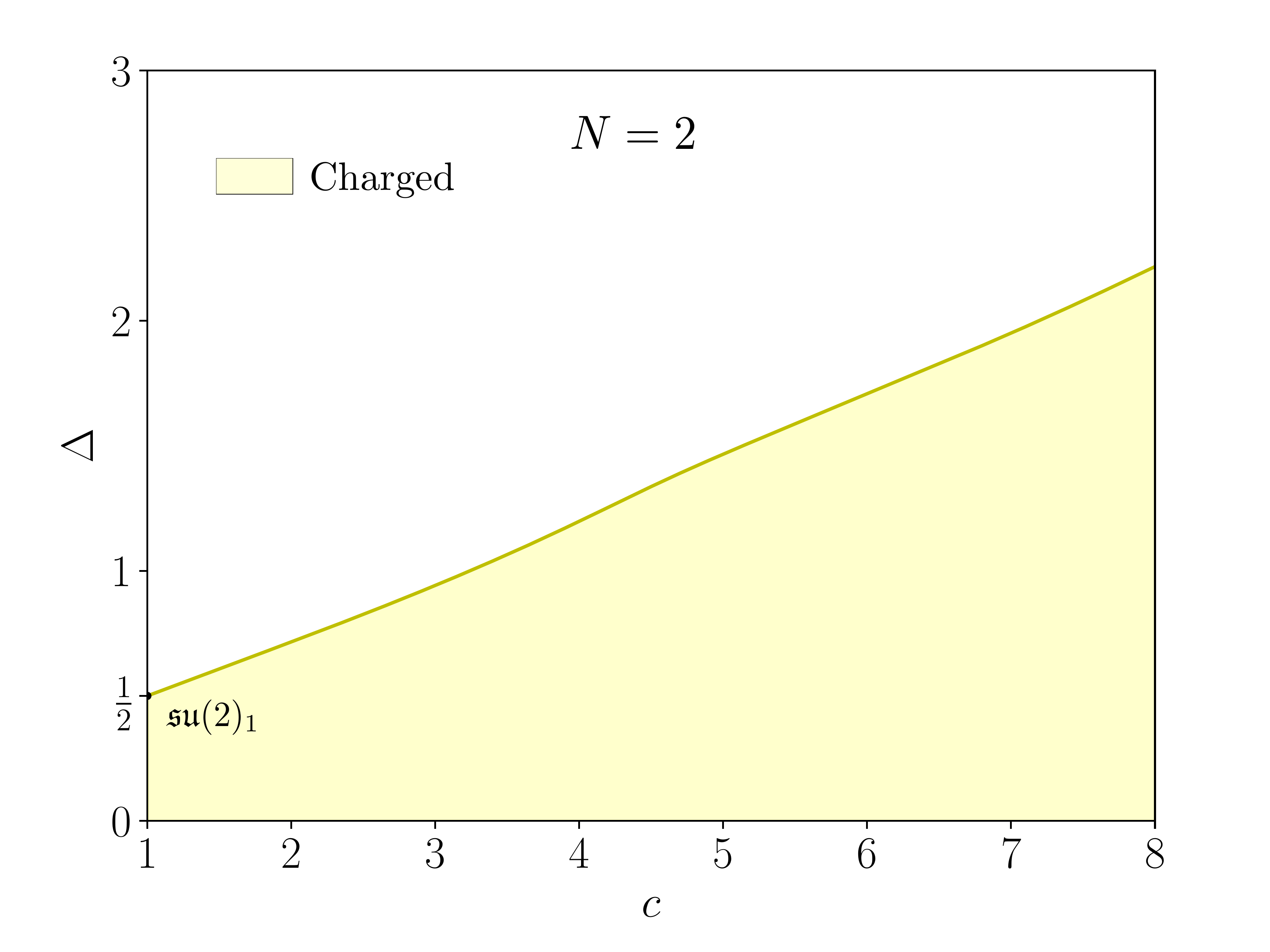}
\includegraphics[width=0.458\textwidth]{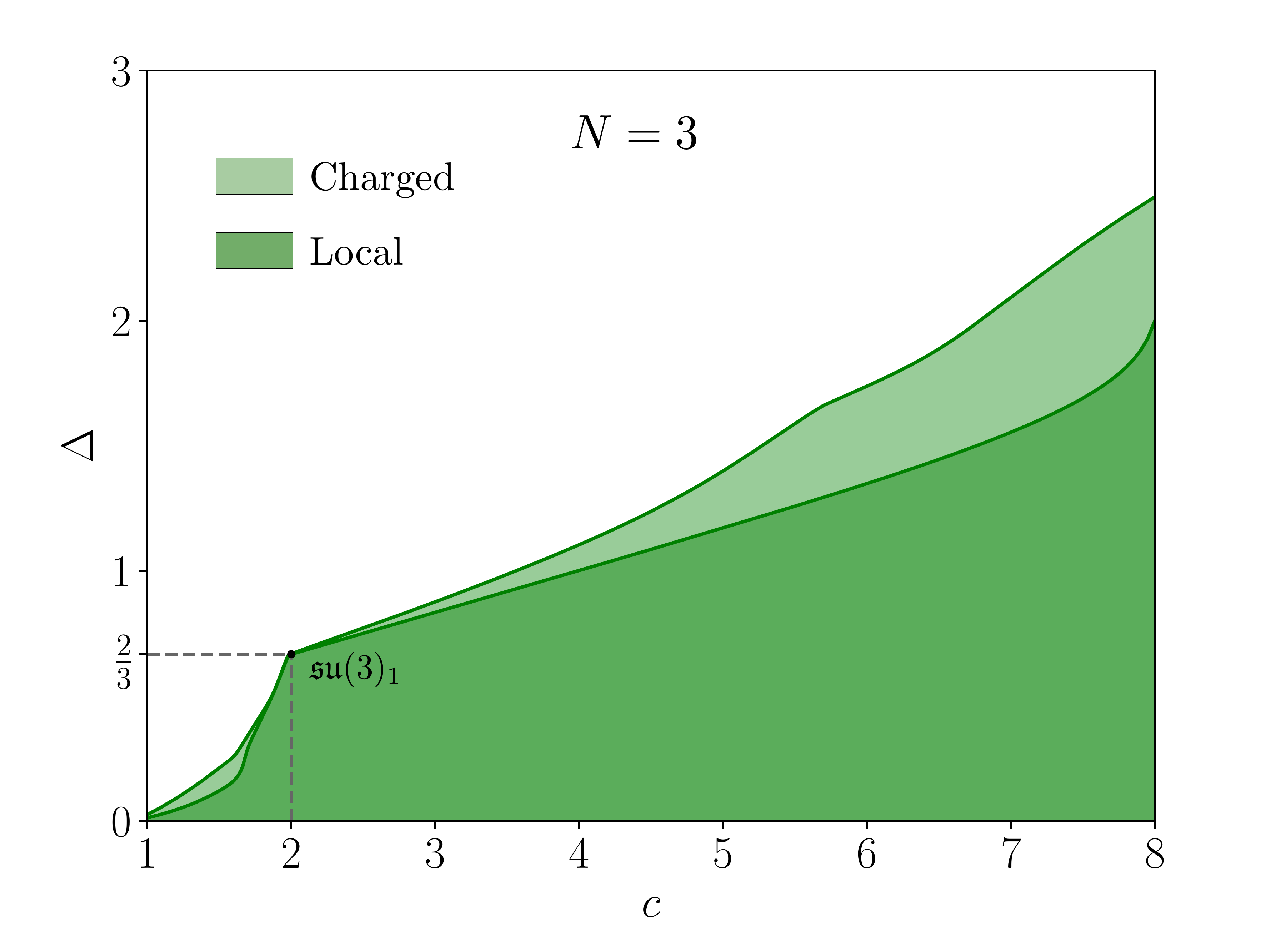}
\includegraphics[width=0.458\textwidth]{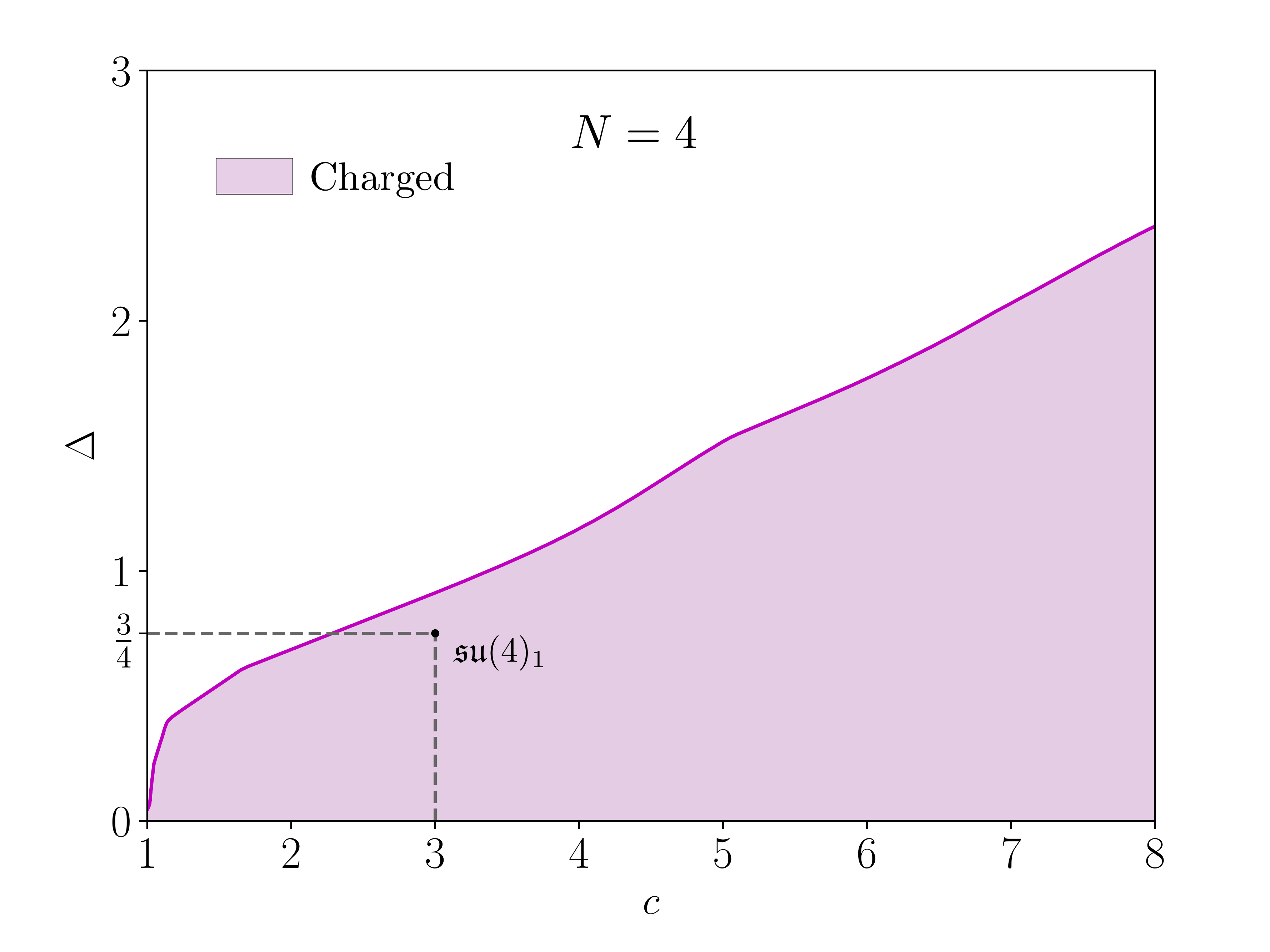}
\includegraphics[width=0.458\textwidth]{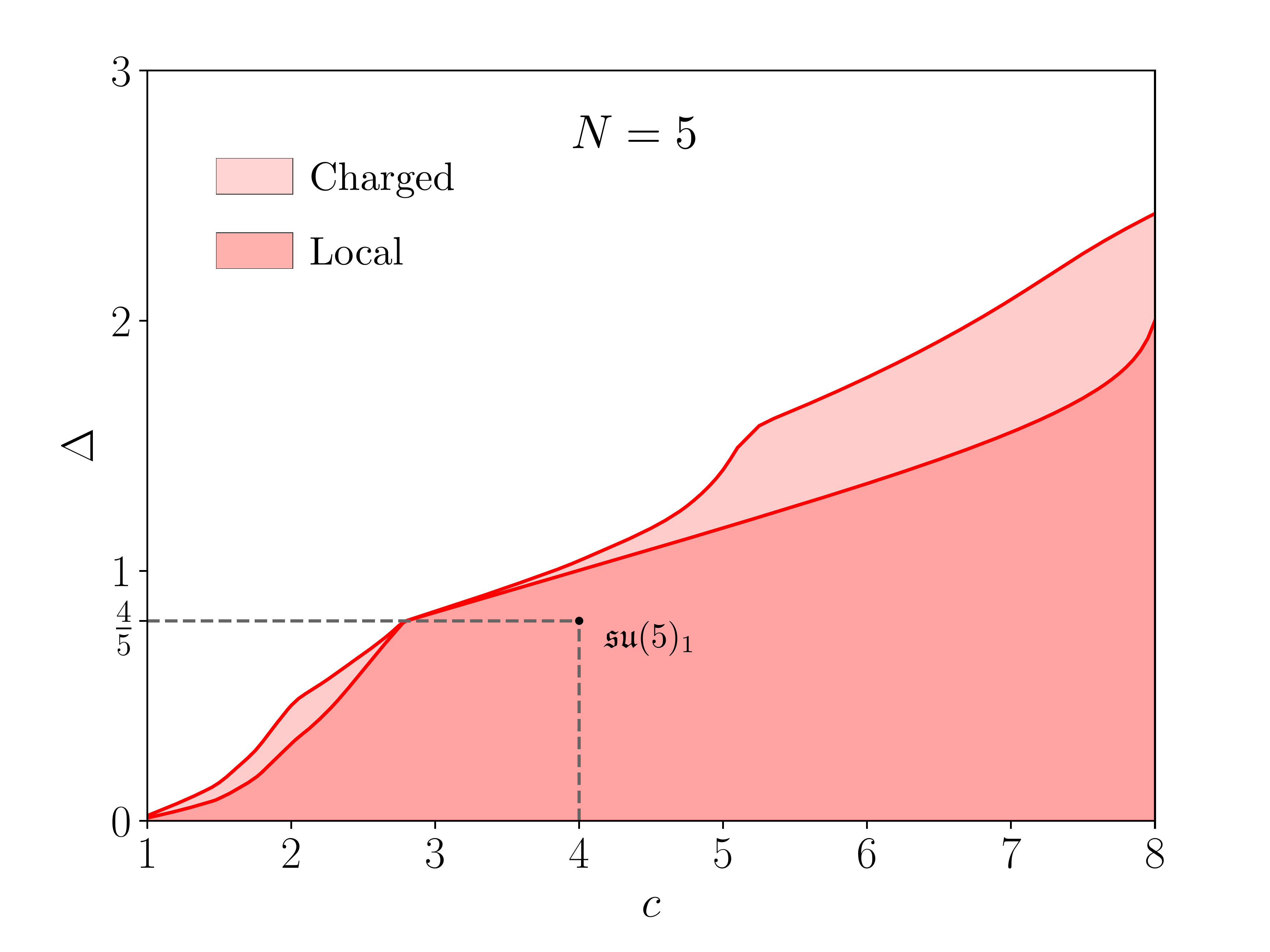}
\includegraphics[width=0.458\textwidth]{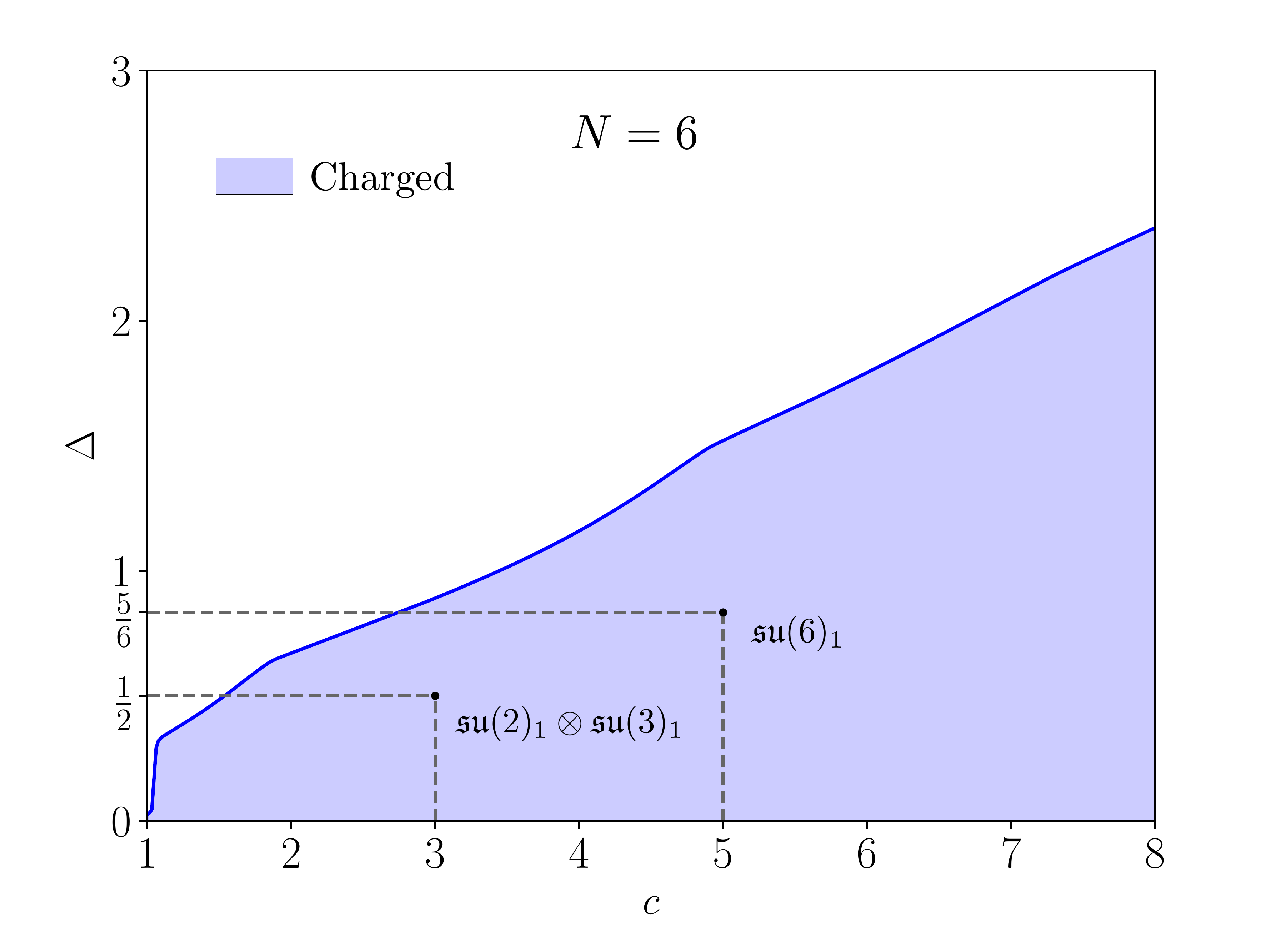}
\caption{Two kinds of universal upper bounds on the scaling dimension of local operators. For $N=2,...,6$, we find an upper bound on the scaling dimension of the lightest charged, scalar operator, where a charged operator is any operator transforming in a non-trivial representation of $\mathbb Z_N^3$. For $N=3,5$, we additionally obtain a stronger bound on the lightest local operator, irrespective of its symmetry properties. The shaded regions represent the allowed regions for the scaling dimensions of such operators in each of the aforementioned cases. The bounds shown here make full use of our improvements to modular bootstrap with global symmetries and anomalies and were computed with $\Lambda^{\text{mod}} = 25$ and $S_{\text{max}}^{\text{mod}} = 50$ (see Section IV for implementation details).}
\label{main_plots}
\end{figure*}
Here we present our main results, which include universal numerical bootstrap bounds on the local primary operator spectrum of unitary, compact (1+1)$d$ CFTs with a $\mathbb Z_N^3$ symmetry and LSM anomaly. We further discuss an application of our numerical bounds to the theory of phase transitions between symmetry protected topological (SPT) phases. The precise details of the LSM anomaly and its implications on the structure of the theories that saturate it will be discussed later. We do not study theories with $c < 1$ since it is known that the unitary models cannot possess the kinds of symmetries, let alone anomalies, that we study in this work\cite{Ruelle:1998zu}. \par
We obtain three types of numerical bounds, each of which will be an upper bound on the scaling dimension of the lightest scalar primary field transforming in some representation of $\mathbb Z_N^3$. The three possibilities we consider for the representations of operators are the trivial representation, any non-trivial representation, or any representation. The first two cases then are upper bounds on the scaling dimension of the lightest symmetric or charged scalar local primary, respectively, and the last case represents an upper bound on the lightest local, scalar operator. We only present bounds on the lightest $\mathbb Z_N^3$-symmetric operator for $N=2,3$, since for larger $N$ the bounds converge very slowly, and the improvements introduced in this work do not improve our ability to guarantee relevant, symmetric, scalar operators. \par
Before discussing our bounds, we mention that examples of CFTs with the properties necessary for our bootstrap bounds to apply have been discussed, with emphasis on their LSM anomalies, in Ref. [\onlinecite{Alavirad:2019iea}]. As mentioned, the main class of examples for theories with the $\mathbb Z_N^3$ LSM anomalies may be found within the conformal manifolds of $N-1$ compact bosons with certain symmetry constraints. However, we also remark that it is possible to find examples of theories with these LSM anomalies with lower central charge than what is suggested in Ref. [\onlinecite{Alavirad:2019iea}]. When $N = \prod_i n_i$ with $n_i$ all coprime integers, CFTs with the $N = \prod_{i} n_i$ LSM anomaly can be constructed by taking the tensor product of theories with the LSM anomaly corresponding to each $n_i$. The reason for this is that in this situation $\mathbb Z_N = \bigoplus_{i} \mathbb Z_{n_i}$. It can be checked straightforwardly that projective representations of $\mathbb Z_N^2$ can be decomposed into tensor products of the projective representations of $\mathbb Z_{n_i}^2$ and further that the other properties implied by the LSM anomaly corresponding to $N$, such as the spin selection rules of defect operators (see section III and Table \ref{spins_table}), are the same. Using, for instance, the WZW models $\mathfrak{su}(n_i)_1$, we can thus construct theories with the $N = \prod_{i} n_i$ LSM anomaly that have central charge $c = \sum_{i} (n_i-1) < N-1$.
\subsection{Universal bounds on lightest \texorpdfstring{$\mathbb Z_N^3$}{ZNxZNxZN}-symmetric scalar}
Here we discuss our bounds on the lightest $\mathbb Z_N^3$-symmetric scalar operator for $N=2,3$, which are shown in Figure \ref{fig:symmetric_plots}. Bounds on symmetric, scalar operators are interesting primarily to determine whether theories whose central charge is within a certain range \emph{cannot} describe stable gapless phases where the microscopically-imposed symmetry is $\mathbb Z_N^3$. In the presence of such a symmetry-preserving relevant operator, an RG flow may be triggered to a nearby phase with the same symmetry, but due to the anomaly this flow will generally lead either to a phase where the symmetry is spontaneously broken, or perhaps the flow will end at a non-trivial CFT fixed point. In either case, the initial CFT is thus unstable. The calculations performed in this section involve only the standard modular bootstrap setup with global symmetries and $\mathbb Z_N$ anomalies of Refs. [\onlinecite{Lin:2019kpn,Lin:2021udi}], with slight modification to include the larger symmetry group. It turns out that our improvements to this setup, which we use in the remainder of our calculations, do not lead to an enlargement of the range of values of central charge where a relevant symmetric scalar is guaranteed. This trend continues for larger $N$, where our methods do not lead to any non-trivial range of values of the central charge for which a relevant, $\mathbb Z_N^3$-symmetric operator is guaranteed. 
\subsubsection{\texorpdfstring{$N=2$}{N=2}} 
For $N=2$, modular bootstrap is somewhat sensitive to the LSM anomaly since it sees that one of the $\mathbb Z_2$ TDLs is anomalous (see i.e. Table \ref{spins_table}). Our bound leads to a range of values of central charge such that any theory in the range with the $N=2$ LSM anomaly must contain a relevant, $\mathbb Z_2^3$-symmetric scalar operator.  The range is approximately $$1 < c < 3.5565$$
Our bound is saturated at $c=1$ by $\mathfrak{su}(2)_1$, whose lightest operator symmetric under $\mathbb Z_2^3$ is exactly marginal with $\Delta = 2$. 
\subsubsection{\texorpdfstring{$N=3$}{N=3}}
In this case, we see that any CFT with a non-anomalous $\mathbb Z_3^3$ symmetry (or, equivalently for the calculations presented here, a $\mathbb Z_3^3$ symmetry with the LSM anomaly) must have a relevant symmetric scalar if its central charge is $c < 2$. The WZW model $\mathfrak{su}(3)_1$ nearly saturates our bound at $c=2$, whose lightest operator $\mathbb Z_3^3$-symmetric is exactly marginal. In Ref. [\onlinecite{Lin:2021udi}], it was shown also that when $c < 2$ a $\mathbb Z_3$-symmetric, relevant, scalar operator must be present. The analogous bounds for $\mathbb Z_3^3$ cannot be stronger than this bound, since imposing more symmetry can only make it harder for a given operator to remain symmetric, so it is interesting that our $\mathbb Z_3^3$ bound is still powerful enough to guarantee relevant operators in this range of central charge.  
\subsection{Universal bounds on lightest \texorpdfstring{$\mathbb Z_N^3$}{ZNxZNxZN}-charged scalar}
We now discuss our bounds on the lightest $\mathbb Z_N^3$ charged operator, which are shown in Figure \ref{main_plots}. We define a $\mathbb Z_N^3$-charged operator as a local operator transforming in any non-trivial representation of $\mathbb Z_N^3$. For odd, prime $N$, this choice loses no generality since all non-trivial representations are equivalent in the sense that they are related by outer automorphisms, i.e. a relabeling of the group elements. This relabeling is allowed since each $\mathbb Z_N$ TDL for a given $N$ has the same spin-selection rule (see section III and Appendix A). For even $N$, there are different classes of representations, so in principle more refined bounds could be obtained by bounding the lightest operator in each class separately but, for simplicity, we will not present such bounds. \par 
\subsubsection{\texorpdfstring{$N=2$}{N=2}}
Our $N=2$ bound does not contain many features. The known theories with minimal central charge and the $N=2$ LSM anomaly are the $c=1$ compact boson theories on the circle branch. The theory that maximizes the gap in the $\mathbb Z_2^3$-charged operator spectrum is the WZW model $\mathfrak{su}(2)_1$, whose lightest charged, scalar operator has scaling dimension $\Delta= \frac{1}{2}$. 
\subsubsection{\texorpdfstring{$N=3$}{N=3}}
Our bound on the lightest charged operator for $N=3$ has various interesting features. First of all, the bound approaches $\Delta = 0$ as $c \to 1$, which is in agreement with the analytical analysis of $c=1$ theories that no such theory can have the LSM anomaly for $N=3$. This is quite striking in modular bootstrap calculations, which typically are not quite so strong, and is a feature shared by our bounds for other choices of $N > 2$. The most obvious feature is that the $\mathfrak{su}(3)_1$ WZW CFT, which is the principal example of a WZW CFT with the $N=3$ LSM anomaly, sits at a prominent kink of our upper bound. The $\mathfrak{su}(3)_1$ theory has $c=2$ and its lightest charged scalar has scaling dimension $\Delta = \frac{2}{3}$. \par 
We stress that in this case, since $N$ is odd, modular bootstrap alone does not give a bound at all on the lightest charged operator since all $\mathbb Z_3$ subgroups, in this case, have no $\mathbb Z_3$ anomaly. 
\subsubsection{\texorpdfstring{$N=4$}{N=4}}
This bound displays a few kink-like features at low values of the central charge that we cannot, at present, explain. The $c=3$ compact boson theories are the only theories we know of with the $N=4$ LSM anomaly, among which is the WZW model $\mathfrak{su}(4)_1$. It is expected that $\mathfrak{su}(4)_1$ maximizes the scalar gap among the toroidal compactification CFTs at $c=3$\cite{Afkhami-Jeddi:2020ezh,Angelinos:2022umf} (Maximizing the scalar gap in the moduli space of such theories at generic integral $c$ is a difficult problem. See additionally Ref. [\onlinecite{Benjamin:2021ygh}] for interesting work related to this.). However, the question of which CFT absolutely maximizes the scalar gap at $c=3$ remains an interesting puzzle; the universal upper bound on the scalar gap calculated in Ref. [\onlinecite{Collier:2016cls}] is not saturated by $\mathfrak{su}(4)_1$ at $c=3$. Consequently, we do not know whether there is any $c=3$ theory that saturates our even larger upper bound on the lightest $\mathbb Z_4^3$ charged operator for theories with the $N=4$ LSM anomaly.  
\subsubsection{\texorpdfstring{$N=5$}{N=5}}
This bound is similar to the bound for $N=3$ since again modular bootstrap alone gives no bound. Below $c=4$, which is the minimal central charge we know of where theories with the $N=5$ LSM anomaly are known to exist, we see two features that resemble kinks. The first occurs at $c \approx 2$ and is somewhat soft. Our upper bound at $c=2$ is approximately equal to $\Delta = 0.462$. At this time we are not aware of any evidence to suggest that this corresponds to an actual theory, so it will be interesting to study this feature in future work. The second kink is more sharply defined and is somewhat more intriguing. The location of this kink, according to our plot in Figure \ref{main_plots} which was computed with derivative orders $\Lambda^{\text{mod}} = \Lambda^{\text{cor}}= 25$ (for an explanation of the parameters describing our computational setup, see section IV), is almost exactly at $c=\tfrac{14}{5}$, where our upper bound is equal to $\Delta = 0.8006$. This puts the WZW model $(\mathfrak g_2)_1$ essentially right at the location of this kink. However, as we discuss in the next subsection, we were able to rule out $(\mathfrak g_2)_1$ from having the $N=5$ LSM anomaly with a more intensive calculation, so the origin of this kink remains a mystery. \par 
At $c=4$, the WZW CFT $\mathfrak{su}(5)_1$ is well within the allowed region, but as was the case with $N=4$ we are not sure of a systematic way to maximize the scalar gap among the $c=4$ compact boson theories, subject to the appropriate symmetry requirements, let alone among all CFTs, to see whether ultimately our bound is saturated by an actual CFT. Using the formalism developed in Refs. [\onlinecite{Dymarsky:2020qom, Angelinos:2022umf}] seems like a promising approach, but we leave this interesting exercise to future work.
\subsubsection{\texorpdfstring{$N=6$}{N=6}}
This bound resembles closely our bound for $N=4$, where we see some unexplained features at relatively low central charge. This is the first case where $N$ is a product of coprime integers, so we can realize theories with the $N=6$ LSM anomaly by taking the tensor product of any $N=2$ theory with a $N=3$ theory. Thus, using our currently known examples, we can produce theories with central charge equal to either $c=3$ or $c=5$ by taking the corresponding number of compact bosons and imposing the appropriate symmetries. Examples of WZW models (or tensor products thereof) in this category include $\mathfrak{su}(2)_1 \otimes \mathfrak{su}(3)_1$ and $\mathfrak{su}(6)_1$. Our analysis here is limited again by the issues we mentioned for $N=4,5$, and the WZW models we mentioned are even farther from achieving saturation with our bounds. 
\subsection{Universal bounds on lightest local scalar}
As we are especially interested in finding numerical evidence for CFTs with the $\mathbb Z_N^3$ LSM anomaly with $c < N-1$ for odd $N$, to which the suggested lower bound in Ref. [\onlinecite{Alavirad:2019iea}] applies, we additionally obtain bounds on the lightest local operator transforming in any representation of $\mathbb Z_N^3$. This is the strongest gap assumption we can make, while still being universal, and thus gives, numerically, the strongest bounds. We thus expect any features corresponding to actual theories to have the most clarity within these bounds. To obtain these bounds, we incorporate the correlator bootstrap lower bounds on central charge from Figure \ref{fig:cent_bounds} in addition to the bounds on scaling dimension from Figure \ref{fig:scaling_bounds}. This leads to a notable improvement of our bounds at relatively low central charge; we note that the stronger gap assumption within modular bootstrap alone did not lead to a significant difference in the resulting bounds. 
\subsubsection{\texorpdfstring{$N=3$}{N=3}}
As expected, the theory $\mathfrak{su}(3)_1$ continues to lie at the $c=2$ kink of this bound, which is sharpened further by the stronger assumptions placed on the operator content. Interestingly, at $c \approx 1.7$ it appears that another feature emerges. However, this feature does not take a sharper shape upon using more derivatives in the linear functional within modular bootstrap, as our bounds are very close to saturation in this range of central charge. The question of whether or not this feature is due to an actual CFT is left open. Above $c=2$, our bound converges to the bound on the lightest scalar operator in any CFT obtained in Ref. [\onlinecite{Collier:2016cls}].
\subsubsection{\texorpdfstring{$N=5$}{N=5}}
The feature present at $c \approx 2$ in the bound on the lightest charged scalar did not survive the stronger assumptions used to obtain this bound. Thus, if that feature is due to an actual CFT with the $N=5$ LSM anomaly, it would have to be symmetric under the $\mathbb Z_5^3$ symmetry that carries the anomaly. \par 
Similarly to the $N=3$ case, the stronger assumptions used to obtain this bound sharpen the kink near $c=2.8$ and, for the size of the functional used in making the $N=5$ plot of Figure \ref{main_plots}, the theory $(\mathfrak g_2)_1$ is not yet ruled out. However, we also did a calculation where we improved some of our parameters to $\Lambda^{\text{mod}}=41$ and $S^{\text{mod}}_{\text{max}} =80$. Our bound at $\Lambda^{\text{mod}} = 25$ is very close to saturation in this range of central charge, but nonetheless we obtained an upper bound of $\Delta = 0.79990$ at $c=2.8$, thus ruling out $(\mathfrak g_2)_1$. Studying this kink further is an interesting task we leave to future work. Even though we have not yet exactly pinned down the location of the kink, it seems a strong possibility that, in this case, we have uncovered evidence for a non-trivial example of some theory with the $N=5$ LSM anomaly that has $c < 4$.
\subsection{Application to multicritical points of (1+1)\texorpdfstring{$d$}{d} symmetry protected topological phases}
Symmetry protected topological (SPT) phases are a special class of gapped Hamiltonians with a global symmetry $G$. When placed on a manifold without boundary, SPT phases possess a unique, $G$-symmetric ground state that cannot be connected to a trivial product state by applying a finite depth local unitary circuit (FDLUC), where the local unitaries in the circuit individually preserve the global symmetry\cite{Chen:2010gda}. In one spatial dimension, SPT phases protected by a symmetry group $G$ are classified by the group cohomology group $H^2(G,U(1))$. The group cohomology group encodes the algebraic structure of the phases under stacking, which is determined by its group multiplication\cite{Chen:2011pg}. Further, the class $[\alpha] \in H^2(G,U(1))$ labelling each phase determines the projective representation carried at each boundary endpoint when a Hamiltonian in a non-trivial SPT phase is placed on a lattice with a boundary. \par 
An interesting area of study is that of the nature of second-order phase transitions between SPTs\cite{Tsui:2015dja,Tsui:2017ryj,Verresen:2017cwj,Tsui:2019piu}. It has been argued by Bultinck\cite{Bultinck:2019zzo} that for any SPT phase $[\alpha]$ satisfying the property that two copies of it is in the trivial SPT phase, i.e. $[\alpha]^2 = [1]$, any critical point describing a second-order phase transition between $[1]$ and $[\alpha]$ will have an emergent $\mathbb Z_2$ symmetry having a mixed anomaly with the internal symmetry $G$ of the neighboring SPT phases. The mixed anomaly is given by a type-III cocycle $\omega \in H^1(\mathbb Z_2, H^2(G,U(1))) \subset H^3(\mathbb Z_2\times G, U(1))$, which has the interpretation, in CFT language, that a defect operator of the emergent $\mathbb Z_2$ symmetry carries a projective representation of $G$. 
\par 
We can roughly sketch the argument for the emergent symmetry and anomaly as follows. A key feature of an SPT Hamiltonian is that there is a FDLUC $\mathcal U$ building its ground state from a trivial product state. Importantly, a FDLUC preserves the correlation length. \par 
Now, consider a one-parameter path of Hamiltonians that passes through a critical point separating the phases $[1]$ and $[\alpha]$ where $[\alpha]^2 = [1]$. We will assume that this critical point is a CFT, and we will restrict the path of Hamiltonians to lie in the vicinity of the critical point such that the low energy spectrum everywhere along the path is reproduced by a Hamiltonian of the form 
\begin{align}
    H(\delta) = H_{\text{CFT}} + \delta \Phi
\end{align}
where $\Phi$ is a perturbation by a relevant operator in the CFT. We expect this scenario since, generically, the perturbation $\Phi$ will open a gap and SPT phases are gapped. Without loss of generality, we assume that $H(\delta)$ is in the trivial phase when $\delta < 0$ and in the non-trivial phase when $\delta > 0$. Next, note that given any Hamiltonian in the trivial SPT phase $H_0$, there exists a $G$-symmetric FDLUC $\mathcal U$ for which $H_\alpha \equiv \mathcal U H_0 \mathcal U^\dagger$ is in the non-trivial SPT phase and, additionally, has an identical correlation length. It is argued that $\mathcal U$ becomes a symmetry at the critical point. At a minimum, we can conclude that $\mathcal U$ is a $\mathbb Z_2$ symmetry, but in some cases it is possible to argue for an even larger emergent symmetry at similar critical points\cite{Tantivasadakarn:2021wdv,Tantivasadakarn:2021aaf}. Thus $\mathcal U$ (really, the restriction of $\mathcal U$ to the low-energy Hilbert space) must precisely be the unitary that changes the sign of the perturbation i.e. $\mathcal U \Phi \mathcal U^\dagger = - \Phi$. Further, by the properties of SPT states, we would expect that upon truncating $\mathcal U$ to an interval, $\mathcal U$ would no longer commute with the global symmetry; instead, the parts of $\mathcal U$ deep in the bulk would commute, but the boundary would host a projective representation of $G$\cite{Else:2014vma}. This should be reflected in the CFT through the 't Hooft anomaly of the full $\mathbb Z_2 \times G$ symmetry, which is captured by a 3-cocycle $\omega \in H^1(\mathbb Z_2,H^2(G,U(1))) \subset H^3(\mathbb Z_2 \times G, U(1))$. In total, at the critical point $\mathcal U$ is an internal symmetry having a mixed anomaly with $G$, and there exists a relevant operator charged under $\mathcal U$ which drives the transition. 

We may now produce a simple generalization of this result to the case of a multicritical point between SPT phases generated by a phase $[\alpha]$ such that $[\alpha]^N = [1]$. Such a family of phases is given, for example, by SPT phases protected by $G = \mathbb Z_N\times Z_N$ for which $H^2(G,U(1)) = \mathbb Z_N$, which is relevant to this work. \par
\begin{figure}[h]
    \centering
    \includegraphics{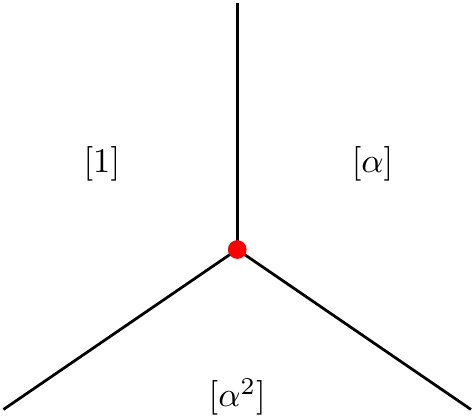}
    \caption{A hypothetical phase diagram where all $N$ of the $\mathbb Z_N\times \mathbb Z_N$ SPT phases, where each phase is labeled by its respective class $[i] \in H^2(\mathbb Z_N\times \mathbb Z_N,U(1)) = \mathbb Z_N$, meet at a multicritical point (marked in red), illustrated in particular for the case $N=3$. The dimension of such a phase diagram increases with $N$. We expect the multicritical point to possess a mixed anomaly of the kind studied in this work, and our numerical bounds restrict the properties of CFTs that can describe such multicritical points.}
    \label{fig:SPT_phase_diagram}
\end{figure}

 
In this setting, we may now consider an $(N-1)$-parameter family of Hamiltonians
\begin{align}
    H(\delta_1,...,\delta_{N-1}) = H_{\text{CFT}} + \sum_i \delta_i \Phi_i
\end{align} 
and essentially repeat the argument above with minimal modification. The main difference between the previous case and this case is that the FDLUC $\mathcal U$ will cyclically permute the $N$ different phases leading to a $\mathbb Z_N$ action among the fields that perturb the CFT into the different SPT phases, i.e. $\mathcal U \Phi_i \mathcal U^\dagger = \sum_j U_{ij} \Phi_j$ with $U^N = I$. We have presented an illustration of such a hypothetical critical point in Figure \ref{fig:SPT_phase_diagram} for the case $N=3$. \par 
The transition to the neighboring SPT phases for such a multicritical point would be driven by relevant perturbations that are charged under the emergent $\mathbb Z_N$ but preserve the microscopic $\mathbb Z_N\times \mathbb Z_N$. An interesting class of theories to consider are ones where there are no relevant operators that are symmetric under the full $\mathbb Z_N^3$ symmetry. This is the generic case that is expected without additional fine-tuning. For $N=3$, we see that such critical points, which do not contain $\mathbb Z_3^3$-symmetric relevant operators, must have $c \ge 2$ using Figure \ref{fig:symmetric_plots}. On the contrary, with a symmetry-preserving relevant perturbation there could be neighboring phases where $\mathcal U$ is an exact lattice symmetry (but, generically, it will not necessarily be a $\mathbb Z_N$ symmetry). Due to the anomaly, these phases cannot be trivially gapped and therefore must be SSB or gapless. \par 
\section{LSM anomalies from topological defect lines}
The microscopic assumptions leading to the generalized LSM theorem manifest in a CFT as a particular set of properties possessed by so-called topological defect lines (TDLs) that implement the symmetries. In this section we describe these properties with emphasis on aspects that enter in bootstrap calculations. 
\subsection{Topological defect lines}
A 0-form global symmetry of an internal symmetry group $G$ in a (1+1)d CFT is given by a linear action of the elements in $G$ on the local fields of the theory in a way that commutes with taking the OPE and acts trivially on the stress tensor. When $G$ is continuous, Noether's theorem implies the existence of some conserved 1-form densities $J_i$ whose integrals over closed contours $\mathcal C$ give rise to operators that generate symmetry transformations on the fields of the theory. This allows for the construction of the unitary operator implementing the action of a symmetry group element $g$ 
\begin{align*}
    U_{\mathcal C} (g) = \mathcal T \exp \left[i \int_{\mathcal C} \! *J_i \epsilon^i\right]
\end{align*}
where the $\epsilon^i$ parameterize $G$. When $\mathcal C$ encloses some number of field insertions in a correlation function, we may replace that configuration with one where each field is replaced by its image under the action of $g$ upon shrinking $\mathcal C$ down to a point.  In particular, the value of the correlation function does not change under smooth deformations of $\mathcal C$ that do not cross any field insertions. Since TDLs commute with the stress-energy tensor, which is the reason for their topological property, encircling a TDL around a local operator leaves its conformal dimensions unchanged. Thus, any two local operators related by an internal symmetry transformation must have identical scaling dimensions and spins.

The notion of a TDL can also be extended to the case when $G$ is discrete. In this instance, we may not always be able to construct the unitary operator implementing the symmetry as an integral of some local density, but nonetheless we can still associate, to each group element, an extended, oriented topological line operator $\mathcal L_g$ that implements the symmetry by analogy to the continuous case. Two TDLs $\mathcal L_g$ and $\mathcal L_h$ supported on topologically equivalent lines may be fused together to create the line $\mathcal L_{gh}$, so the fusion of TDLs exactly obeys the group multiplication law of $G$. For a given TDL $\mathcal L$, we will denote its orientation reversal (which corresponds to the inverse line when $\mathcal L$ is invertible) by $\bar{\mathcal L}$. We will denote a unitary operator acting on the Hilbert space $\mathcal H$ of local operators, i.e. a TDL $\mathcal L_g$ supported on a time slice, by $\hat{\mathcal L}_g$, and we will denote the action of such operators on a field $\phi$ as $\hat {\mathcal L}_g \cdot \phi$. TDLs have more explicit constructions as well, which can be studied by identifying topological boundary conditions via the folding trick. TDLs are in one-to-one correspondence with such topological interfaces.  The mathematical structure that encodes global symmetries, as well as their `t Hooft anomalies, in a (1+1)d CFT is known as a fusion category. This theory is developed nicely by Chang et. al. in Ref. [\onlinecite{Chang:2018iay}], so we invite the interested reader to consult their work, and mathematical literature referenced therein, for the complete picture.\par
\begin{figure}[h]
    \centering
    \includegraphics{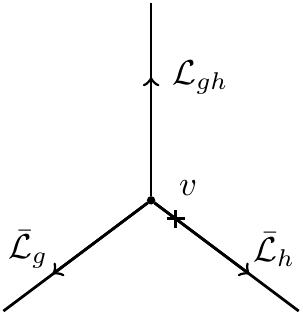}
    \caption{A 3-way junction with junction vector $v \in V_{\mathcal L_{gh},\bar{\mathcal L}_g, \bar{\mathcal L}_h} $. The convention for the junctions is that we label lines oriented pointing outwards from the junction.}
    \label{fig:3-way-junction}
\end{figure}

 
A given TDL $\mathcal L_{g}$ may terminate at a point, and at this point will live a \emph{defect operator}. Such an operator may be constructed through the operator state mapping, since quantizing the theory in a cylinder geometry with a boundary condition twisted by $\mathcal L_{g}$ is equivalent to the theory in radial quantization with a primary field at the origin attached to a TDL. The possible $\mathcal L_g$ defect operators, which we will as $\phi^g$, live in the \emph{defect Hilbert space} $\mathcal H_{\mathcal L_g}$ (sometimes we will refer to the defect Hilbert spaces as twisted sectors). More generally, multiple TDLs $\mathcal L_{g_1},...,\mathcal L_{g_n}$ may terminate at a $n$-way junction and correspondingly there will be a \emph{junction Hilbert space} $\mathcal H_{\mathcal L_{g_1},...,\mathcal L_{g_n}}$,  defined similarly. Due to the fusion properties of the lines, there is an isomorphism between $\mathcal H_{\mathcal L_{g_1,...,g_n}}$ and $\mathcal H_{\mathcal L_{g_1},...,\mathcal L_{g_n}}$. The junction Hilbert space may contain a state which is proportional to the vacuum of the bulk theory with conformal dimensions $(0,0)$, which is unique for the CFTs we consider in this work. We denote the space of such states in $\mathcal H_{\mathcal L_{g_1},...,\mathcal L_{g_n}}$ by $V_{\mathcal L_{g_1},...,\mathcal L_{g_n}}$ and refer to it as the \emph{junction vector space}. Since the TDLs we consider are all invertible, their fusion does not involve direct summation so each $V_{\mathcal L_{g_1},...,\mathcal L_{g_n}}$ is either one or zero dimensional depending on whether the TDLs forming the junction fuse to the trivial line or not. Thus, there is a unique \emph{junction vector} $v \in V_{\mathcal L_{g_1},...,\mathcal L_{g_n}}$ for each junction which keeps track of the overall phase associated with each junction, relative to the bulk vacuum state. In Figure \ref{fig:3-way-junction} we illustrate a 3-way junction, where the ``$\times$'' keeps track of the cyclic ordering of the outward-pointing TDLs. 

\begin{figure}[h]
    \centering
    \includegraphics[]{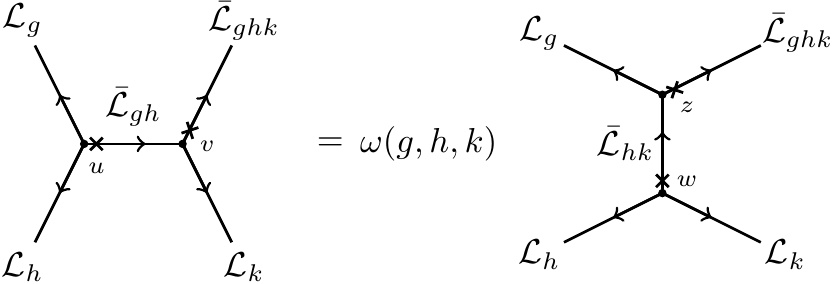}
    \caption{For any configuration of invertible TDLs involving external TDLs $g,h,k$ and $(ghk)^{-1}$, there are two ways to resolve the configuration into one involving only 3-way junctions. The two decompositions differ by a 3-cocycle phase $\omega$.}
    \label{fig:f_move}
\end{figure}

In a theory with a certain set of TDLs, it is possible to decorate the spacetime manifold that the theory lives on with a network of TDLs. This network will consist of some arrangement of lines connected at $n$-way junctions. Using the locality property outlined in Ref. [\onlinecite{Chang:2018iay}], it is possible to resolve any such network into an equivalent one involving only 3-way junctions. However, the resolution of a given network into 3-way junctions is not unique, and there may be a phase difference between different resolutions that cannot be removed by a change of basis in the junction vector spaces. These phase ambiguities are the signal of an 't Hooft anomaly of the $G$ symmetry in the TDL formalism. The possible physically distinct `t Hooft anomalies of a discrete symmetry $G$ in a (1+1)$d$ CFT are classified by group cohomology classes $[\omega] \in H^3(G,U(1))$. Choosing some representative 3-cocycle $\omega(g,h,k)$ from $[\omega]$ allows for the calculation of the phase difference associated with performing a rearrangement of TDLs, sometimes known as an F-move, within some larger network of TDLs, as shown in Figure \ref{fig:f_move}. \par
\begin{figure}[h]
    \centering
    \includegraphics{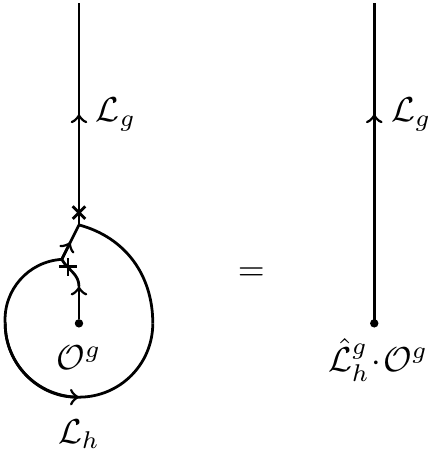}
    \caption{In each defect Hilbert space $\mathcal H_g$ the unitary $\hat{\mathcal L}^g_h$ implementing the action of an element $h \in G$ can be defined by wrapping an $h$-TDL counter-clockwise around a defect operator $\mathcal O^g$ inserted at the origin (in radial quantization) and collapsing the loop down to a point, as shown. This defines the action $\hat{\mathcal L}^g_h \ket{\mathcal O^g}$ on states in $\mathcal H_g$. We will take $U \cdot \mathcal O \equiv U \mathcal O U^\dagger$ to mean conjugation of $\mathcal O$ by the unitary $U$. One can also perform an $F$-move on this configuration to obtain a different convention for resolving the symmetry action, which would differ possibly from this one by a phase for invertible TDLs. Note that if $g$ is the trivial line this reduces to the usual way that symmetry acts on local operators.}
    \label{fig:defect_lasso}
\end{figure}

 
The effect of 't Hooft anomalies in this context is to extend the possible symmetry representations of defect operators to include fractionalized representations. Specifically, we may construct unitary operators $\hat{\mathcal L}_h^g$ that implement the $h$ symmetry in $\mathcal H_g$ for any $g, h \in G$. The TDL configuration defining this unitary is shown in Figure \ref{fig:defect_lasso}. To determine aspects of the symmetry action in the defect Hilbert spaces, one may perform a sequence of F-moves to compare $\hat{\mathcal L}_h^g \hat {\mathcal L}_k^g$ with $\hat{\mathcal L}^g_{hk}$, which will be equal up to a phase i.e.
\begin{align}
    \hat{\mathcal L}_h^g \hat{\mathcal L}_k^g = \chi^g(h,k) \hat{\mathcal L}^g_{hk}. \label{proj_rep}
\end{align}
The phases $\chi^g(h,k)$ are related to the slant products of $[\omega]$, which are given by the representative
\begin{align}
    \chi^g(h,k)=\frac{\omega(g,h,k) \omega(h,k,g)}{\omega(h,g,k)}. \label{slant}
\end{align}
When calculated with the TDL formalism, the representation of $\chi^g(h,k)$ is more complicated, but the essential information is captured by the cohomology class of $\chi^g$, for which it is sufficient to use (\ref{slant}). Note that to correctly reproduce the classification of the symmetry properties of defect operators, when $G$ is Abelian the cohomology class $[\chi_g]$ should be viewed as an element of $H^2(G,\mathbb Z_M)$ when $g$ generates a $\mathbb Z_M$ subgroup. This is roughly the statement that $M $ copies of a $\mathbb Z_M$ defect operator should transform as a local operator, and $H^2(G,\mathbb Z_M)$ classifies the different ways in which this is possible. The symmetry properties of defect operators under the full internal symmetry group of the theory can be thought of as topological invariants, which lead to a finer classification of CFTs enriched by symmetries. For recent explorations of this idea, see e.g. Ref. [\onlinecite{Verresen:2019igf}].\par 
One class of `t Hooft anomalies that have been studied in recent modular bootstrap work involve a single $\mathbb Z_N$.  These so-called type-I anomalies manifest as anomalous spin-selection rules in the defect Hilbert spaces, which can be directly used as input into the modular bootstrap \cite{Lin:2019kpn, Lin:2021udi}.  In contrast, we will be mostly interested in type-III anomalies\cite{deWildPropitius:1995cf}––in particular, the LSM anomaly––whose signatures are more subtle, involving defect operators in the defect Hilbert spaces of one $\mathbb Z_N$ subgroup transforming in a non-trivial projective representation of the remaining $\mathbb Z_N^2$ subgroup. This fact can be inferred directly from the microscopic assumptions in the lattice setting. If the lattice system initially transforms in a linaer representation of $G$, adding a site, i.e. adding a translation symmetry defect, forces the system to realize $G$ projectively.  \par
\begin{table}[]
{\renewcommand{\arraystretch}{1.2}%
\begin{tabular}{c@{\hspace{0.4cm}}c@{\hspace{0.4cm}}c@{\hspace{0.4cm}}c@{\hspace{0.4cm}}c@{\hspace{0.4cm}}c@{}}
                       & \multicolumn{5}{c}{Order of $g \in G$}                                                                              \\ \cmidrule(l){2-6} 
$N$                    & 2                  & 3                  & 4                  & 5                  & 6                  \\ \midrule
\multirow{2}{*}{$2\,$} & 1 if $g=(1,1,1)$   & \multirow{2}{*}{$-$} & \multirow{2}{*}{$-$} & \multirow{2}{*}{$-$} & \multirow{2}{*}{$-$} \\
                       & 0 else             &                    &                    &                    &                    \\ \midrule
$3\,$                  & $-$                  & 0                  & $-$                  & $-$                  & $-$                  \\ \midrule
\multirow{2}{*}{$4\,$} & \multirow{2}{*}{0} & \multirow{2}{*}{$-$} & 2 if $g_i$ odd     & \multirow{2}{*}{$-$} & \multirow{2}{*}{$-$} \\
                       &                    &                    & 0 else             &                    &                    \\ \midrule
$5\,$                  & $-$                  & $-$                  & $-$                  & 0                  & $-$                  \\ \midrule
\multirow{2}{*}{$6\,$} & 1 if $g = (3,3,3)$ & \multirow{2}{*}{0} & \multirow{2}{*}{$-$} & \multirow{2}{*}{$-$} & 3 if $g_i$ odd     \\
                       & 0 else             &                    &                    &                    & 0 else             \\ \bottomrule
\end{tabular}}
\caption{The values $k$ of $\mathbb Z_M$ anomalies appearing in (\ref{zn_anom}) for elements of the groups $\mathbb Z_N^3$ for theories with the LSM anomaly, i.e. when the full cocycle is given by (\ref{lsm_cocycle}). The columns are organized by the order of the elements $g=(g_1,g_2,g_3) \in \mathbb Z_N^3$, and the rows each correspond to a different choice of $N$.}
\label{spins_table}
\end{table}
We will now summarize the details relevant to bootstrap of each of these two classes of anomalies, each of which generally appears in the theories we consider.

\subsubsection{\texorpdfstring{$\mathbb Z_M$}{ZM} anomalies}
In the case that $G$ has a $\mathbb Z_M$ subgroup, the subgroup may have a $\mathbb Z_M$ anomaly. In particular, the restriction of the full cocycle, which encodes the full anomaly of $G$, to this subgroup may be cohomologous to
\begin{align}
    \omega(a,b,c) = \exp{\left(2\pi i \frac{ka}{M^2}\left(b+c -  [b+c]_M\right)\right)} \label{zn_anom}
\end{align}
where $a,b,c \in \mathbb Z_M$ and $[k] \in \mathbb Z_M = H^3({\mathbb Z}_M, U(1))$ labels the anomaly. Note that $[a+b]_M \equiv a+b \mod M$. The main signature of a TDL $\mathcal L_g$ generating a $\mathbb Z_M$ symmetry with anomaly $[k]$ is that that the defect Hilbert space $\mathcal H_g$ will contain states whose spins are in \cite{Lin:2019kpn}
\begin{align}
    s \in k/M^2 + \mathbb Z/M \label{spin-select}
\end{align}
Non-zero $k$ thus corresponds to an anomalous spin selection rule for the defect operators. We will refer to a TDL generating a $\mathbb Z_M$ subgroup as non-anomalous if $k=0$ and anomalous otherwise. The anomaly forces the minimum possible scaling dimension for a defect operator on such a TDL to be $\Delta = \min(\tfrac{|k|}{M^2}, \tfrac{|k-M|}{M^2}) $ as opposed to $\Delta = 0$ for a non-anomalous TDL. \par 
A $\mathbb Z_M$ anomaly can be detected by, for instance, studying the torus partition function of the theory twisted by symmetry defects and imposing covariance of the partition function under the modular $S$ transformation, see e.g  Refs. [\onlinecite{Lin:2019kpn,Lin:2021udi}]. To deduce the spin selection rule of a TDL $\mathcal L_g$, with $g$ generating a $\mathbb Z_M$ subgroup, given any 3-cocycle $\omega$, we can use the formula 
\begin{align}
    e^{2\pi i M s} = \prod_{i = 1}^{M-1} \omega(g^{-1},g^i,g)
\end{align}
given in Ref. [\onlinecite{PhysRevD.101.106021}] and solve for the allowed values of $s$. The $\mathbb Z_M$ anomalies possessed by theories with the LSM anomalies that we study are listed in Table \ref{spins_table}. 
\subsubsection{\texorpdfstring{$\mathbb Z_N^3$}{ZNxZNxZN} LSM anomalies}
It is also possible to have mixed anomalies between different $\mathbb Z_N$ subgroups. When three $\mathbb Z_N$ subgroups are involved, it is possible to have the LSM anomaly, the effects of which can be derived using the following representative cocycle $[\omega] \in H^1(\mathbb Z_N, H^2(\mathbb Z_N^2,U(1)))$
\begin{align}
    \omega(f,g,h) = \exp\left(\frac{2\pi i}{N}  f_1 g_2 h_3\right).
    \label{lsm_cocycle}
\end{align}
Here we denote $\mathbb Z_N^3$ elements by i.e. $g = (g_1,g_2,g_3)$. Note that the multiplication on the right-hand side should be viewed as multiplication of real numbers, not the group multiplication for $\mathbb Z_N$ elements. The signature of this anomaly is that defect operators of one $\mathbb Z_N$ subgroup transform in a projective representation of the remaining $\mathbb Z_N\times \mathbb Z_N$ subgroup. \par 
More specifically, let $g_1,g_2$, and $g_3$ be the generators of the three $\mathbb Z_N$ subgroups in $\mathbb Z_N^3$.  Let us now consider a defect operator that lives at the end of a TDL $\mathcal L_{g_1}$. One can show that, in the presence of $\mathcal L_{g_1}$, the TDLs $\mathcal L_{g_2}$and $\mathcal L_{g_3}$ no longer commute when the $\mathbb Z_N^3$ symmetry has this LSM anomaly. Instead, the  unitaries $\hat{\mathcal L}_{g_2}^{g_1}$ and $\hat{\mathcal L}_{g_3}^{g_1}$ satisfy 
\begin{align}
   \hat{\mathcal L}_{g_2}^{g_1}\hat{\mathcal L}_{g_3}^{g_1} = e^{2\pi i /N} \hat{\mathcal L}_{g_3}^{g_1}\hat{\mathcal L}_{g_2}^{g_1}
\end{align}
meaning $g_2,g_3$ are realized projectively. Thus, the entire $g_1$ defect Hilbert space decomposes as a direct sum of such projective representations.
A convenient basis for such $\mathbb Z_N \times \mathbb Z_N$ projective representations is the clock and shift basis. In this basis, $g_2$ is diagonalized so that states in $\mathcal H_{g_1}$ carry definite $g_2$ charge. Then, $g_3$ acts as a shift operator which increases $g_2$ charge by one modulo $N$. That is, we can choose primary fields $\phi_0^{g_1}, ... , \phi_{N-1}^{g_1}$ in $\mathcal H_{g_1}$, all with identical conformal dimensions, such that
\begin{align}
    \hat{\mathcal L}_{g_2}^{g_1} \cdot \phi^{g_1}_I &= e^{2\pi i I/N} \phi^{g_1}_I \label{proj_rep1}\\
    \hat{\mathcal L}_{g_3}^{g_1} \cdot \phi^{g_1}_I &= \phi^{g_1}_{[I+1]_N}.
    \label{proj_rep2}
\end{align} \par 
From this, we see that given a single primary field in $\mathcal H_{g_1}$ we can infer the existence of $N-1$ more fields of the same conformal dimensions.  Additionally, since a unitary CFT has a well-defined Hermitian conjugate, when $g_1 \ne g_1^{-1}$ there also must exist $N$ more fields living in $\mathcal H_{g_1^{-1}}$. Thus, we see that a feature of theories with the $\mathbb Z_N^3$ LSM anomaly is that the degeneracy of Virasoro defect primary fields with fixed conformal dimensions is a multiple of $N$ when the defect corresponds to an order $N$ element of $G$.
\subsection{Gauging non-anomalous TDLs}
A CFT with $G=\mathbb Z_N^3$ and the LSM anomaly described above contains various non-anomalous TDLs which, consequently, may be gauged (in CFT terminology, orbifolded). For the parts of our calculations involving correlation functions of defect operators, it will be somewhat simpler to view things from the point of view of the gauged theory since, from this perspective, many of the technicalities one needs to be aware of when working with defect operators can be avoided\footnote{We thank Sahand Seifnashri and especially Shu-Heng Shao for helpful conversations and suggestions related to the content in this section.}. \par 
Let $H = \mathbb Z_N$ be such a non-anomalous $\mathbb Z_N$ subgroup. Given some CFT $\mathcal T_{\text{LSM}}$ with the LSM anomaly, when we gauge $H$ we end up with a new theory $\tilde{\mathcal T}_{\text{LSM}} = \mathcal T_{\text{LSM}}/H$, which generally will have a different spectrum of local operators. The gauging procedure amounts to first throwing away all local operators that are charged under $H$, leaving only the $H$-invariant operators. This theory, keeping just symmetric local operators, will generically not be modular-invariant, so in addition to the $H$-invariant local operators we must also bring down $H$-invariant defect operators from the $H$-twisted sectors. Due to the spin-charge relation for defect operators\cite{Lin:2019kpn,Lin:2021udi}, this amounts to adding all integer spin defect operators from the $H$-twisted sectors. Since $H^2(\mathbb Z_N,U(1))$ is trivial, we do not additionally need to specify a discrete torsion class as another ingredient to this gauging procedure. \par 
When a set of TDLs is gauged, the gauged theory will, in general, possess a different set of TDLs from the ungauged theory. We may conclude various facts about the TDLs of $\tilde{\mathcal T}_{\text{LSM}}$ with the help of some results contained, for instance, in Ref. [\onlinecite{Bhardwaj:2017xup}] and also Ref. [\onlinecite{Thorngren:2019iar}]. The key thing to note is that local operators must transform in \emph{linear} representations, but, at first sight, the defect operators coming from $\mathcal T_{\text{LSM}}$, which are promoted to local operators of $\tilde{\mathcal T}_{\text{LSM}}$, transform in projective representations of the $G/H = \mathbb Z_N^2$ symmetry. Thus, to restore a linear representation of this symmetry in $\tilde{\mathcal T}_{\text{LSM}}$, the projective phases $\chi^g(h,k)$ for $g \in H$, $h,k \in G/H$ are promoted into central elements of the symmetry group $\tilde G$ of $\tilde{\mathcal T}_{\text{LSM}}$, so we can identify $\tilde G$ as a central extension of $G/H$ by $H$ with presentation
\begin{align}
\langle a,b,c | a^N,b^N,c^N,aba^{-1}b^{-1}c^{-1},aca^{-1}c^{-1},bcb^{-1}c^{-1}\rangle. \label{presentation}
\end{align}
In $\tilde{\mathcal T}_{\text{LSM}}$, the operators transforming in one-dimensional irreducible representations (irreps) of $\tilde G$ were local operators in $\mathcal T_{\text{LSM}}$, and operators transforming in higher dimensional irreps were defect operators of $\mathcal T_{\text{LSM}}$. \par 
When $N$ is prime, the groups with presentation given by (\ref{presentation}) are known as Heisenberg groups $\text{He}(3,\mathbb Z_N)$. When $N = 2$, we can also identify $ \text{He}(3,\mathbb Z_2) \cong D_8$ where $D_8$ is the square dihedral group. This particular case of gauging was studied in a rather different context in Ref. [\onlinecite{Gaiotto:2017yup}]. When $N$ is not prime, the representation theory of the groups that we might still call $\text{He}(3, \mathbb Z_N)$ differs from the prime case. In mathematics literature such groups have a different naming convention, but we will still refer to groups with presentation (\ref{presentation}) as $\text{He}(3,N)$. \par 
We imported information about the representation theory of the groups $\text{He}(3,N)$ with the help of \texttt{GAP}\cite{GAP4}. For each $N$ we consider, the information about the corresponding group is contained in the \texttt{SmallGroups} library. The \texttt{SmallGroups} identification numbers for the groups are (8,3),  (27,3), (64,18), (125,3), and (216,77) for $N = 2,...,6$. We will label representations of $\text{He}(3,N)$ by $R_n$ where $n$ is the index of the representation in \texttt{GAP}. 
\subsection{Comments on correlation functions of defect operators}
The OPE is a mapping between two separated local fields $\phi_a,\phi_b$ and linear combinations of local fields at a single point, schematically expressed as
\begin{align*}
    \phi_a \times \phi_b \sim \sum_{\mathcal O} \lambda_{\phi_a \phi_b\bar{\mathcal O}}\mathcal O
\end{align*}
where the (possibly complex) numbers $\lambda_{\phi_a \phi_b\bar{\mathcal O}}$ are called the OPE coefficients, and $\bar{\mathcal O}$ denotes the Hermitian conjugate of $\mathcal O$. \par 
The OPE can also be taken between defect operators, but it has a slightly different interpretation, since in general the inputs and outputs of the defect OPE will live in different defect Hilbert spaces. Again schematically, if we take defect operators living at the ends of TDLs $\mathcal L_g$ and $\mathcal L_h$, their OPE can be written as follows
\begin{align*}
    \phi^g_a \times \phi^h_b \sim \sum_{\mathcal O^{gh}} \lambda^v_{\phi_a^g \phi_b^h \bar{\mathcal O}^{\overline{gh}}} \mathcal O^{gh}
\end{align*}
where we use the notation $\bar g \equiv g^{-1}$. It is important to note that OPE coefficients of defect operators depend, up to a phase, on a junction vector $v$. This is because in order to calculate the OPE coefficient, one constructs a four-point function containing three defect operators together with a junction vector connecting the TDLs of the defect operators in a configuration such as that of Figure \ref{fig:3-way-junction}, with defect operators added to ends of the TDLs. Further, since there are three different junction vectors for a given 3-way junction (when the TDLs forming the junction are all invertible), there is freedom to choose a junction vector when writing down OPE coefficients. Thus, there is a separate OPE coefficient for each junction vector, but all choices are related to each other in a fixed way via the cyclic permutation map between junction vector spaces, which results in an overall phase difference. Strictly speaking, the OPE coefficient for defect operators may also depend on the angle of the TDL leaving the defect operators, but since we deal only with scalar defect operators, which transform trivially under rotations, we do not encounter this complication.
\begin{figure}[h]
    \centering
    \includegraphics[width=0.35\textwidth]{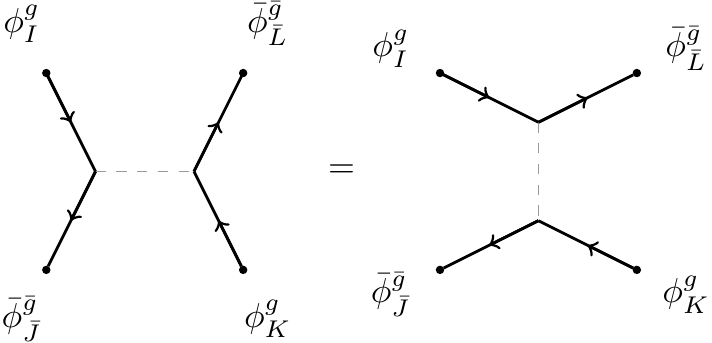}
    \includegraphics[width=0.35\textwidth]{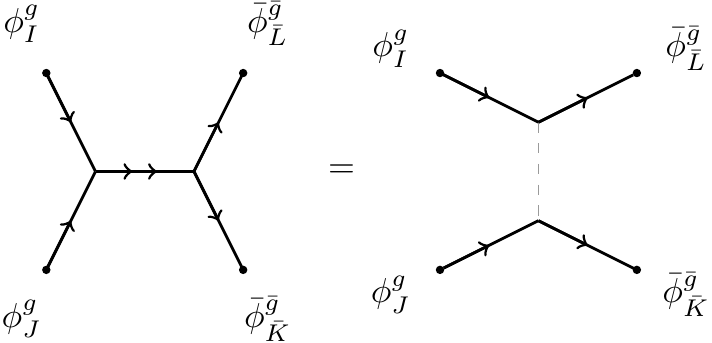}
    \includegraphics[width=0.35\textwidth]{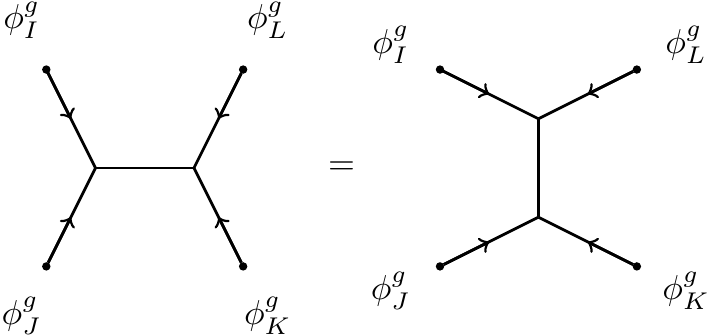}
    \caption{A pictorial representation of the four-point functions of defect operators that we impose crossing symmetry on in our bootstrap calculations. In these configurations, all the external TDLs are either $\mathcal L_g$ or $\bar{\mathcal L}_g$. The internal TDL marked by a double arrow represents $\mathcal L_{g^2}$, but when $\mathcal L_{g^2}$ squares to the trivial line, like when $N=4$, we denote it by an orientation-less TDL such as in the bottom diagram. When $N=2$ all diagrams are equivalent. For general $N > 2$, only the top and middle diagrams are consistent with the group multiplication except when $N=4$ where the bottom diagram is consistent as well. Due to the gauging argument, we can neglect the dependence on the junction vectors since there is a canonical choice that removes the phase $\omega(g,h,k)$ in our cases, which involve only external defect operators hosted on a single $\mathbb Z_N$ TDL or its inverse. We could also even consider these to be correlation functions of local operators, but to emphasize the more generic scenario we stick to drawing defect operators.}
    \label{fig:four_point_funcs}
\end{figure}
\par 
In the most general setting, there are a few differences that must be accounted for when considering correlation functions of defect operators. First of all, the non-local nature of the defect operators leads, in some cases, certain correlation functions of defect operators to be multi-valued maps of the positions of the operators in the correlation function. This is because winding an operator charged under the TDL attached to a defect operator must involve acting on the charged operator with the TDL. Additionally, defect operators obey a modified version of crossing symmetry. A correlation function of four defect operators is really a six-point function of the defect operators along with two junction vectors, which can be represented graphically by attaching defect operators to the TDLs in the left-hand side of the graphical equation in Figure \ref{fig:f_move}. Crossing symmetry is then modified since there may be a phase $\omega(g,h,k)$ accrued when relating the two OPE channels of the four external operators. Further, the intermediate operators will live in potentially different defect Hilbert spaces. For more discussion on correlation functions of defect operators, see also Ref. [\onlinecite{Chang:2020imq,Huang:2021zvu}].
\par Appealing to the gauging argument from the previous subsection, we will now argue that the version of crossing symmetry obeyed by the scalar defect operators we consider in this work is identical to certain local operators. Again, let $\mathcal T_{\text{LSM}}$ be defined as before and $H=\mathbb Z_N$ be a non-anomalous subgroup of $G = \mathbb Z_N^3$. Consider external defect operators $\phi^g_I,\bar{\phi}^{\bar g}_{\bar I}$, for $I = 1...,N$, where we take $g$ to generate $H$. Since scalar defect operators are neutral under $H$, upon gauging $H$ they are promoted to local operators in $\tilde{\mathcal T}_{\text{LSM}}$ that, without loss of generality, can be taken to transform in representations $R_{N^2+1}, \bar R_{N^2+1}$ of $\text{He}(3,N)$, which are $N$-dimensional irreps. Since all operators appearing in OPEs of the defect operators $\phi^g_I,\bar{\phi}^{\bar g}_{\bar I}$ will also be neutral under $H$ by $\mathbb Z_N$ charge conservation, gauging does not modify the values of the correlation functions that they generate. In particular, this means that it must be possible to set the extra phase that could appear in Figure (\ref{fig:f_move}) equal to 1 since we can exactly relate the value of the defect operator correlation function to a correlation function involving only local operators via gauging. Thus, for the purpose of bootstrap, the crossing equations we generate from viewing the external operators as defect operators in $\mathcal T_{\text{LSM}}$ or local operators in $\tilde{\mathcal T}_{\text{LSM}}$ will be identical. We should note, however, that if one wants to strictly think of everything in terms of defect operators, an analogous statement is that all the phases $\omega(\bar g, \bar h, \bar k)$ can be set equal to 1 for the four-point functions we consider, which only involve TDLs corresponding to elements of $H$, by a careful choice of junction vectors, which is done partially in Appendix A of Ref. [\onlinecite{Chang:2018iay}]. We note that, for these reasons, we may drop all dependence on junction vectors from the diagrams we draw and formulae we write down. This is reflected in the diagrams representing the four-point functions we consider, as in figure \ref{fig:four_point_funcs}.\par 
\section{Numerical Bootstrap approach}
The bounds presented in the main results section are obtained via a new approach that uses correlator bootstrap bounds as additional input to modular bootstrap, thereby augmenting the usual procedure to incorporate global symmetries into modular bootstrap. Here we outline the details of how our bounds are obtained.

\subsection{Correlator bootstrap with defect operators}
We first will show both how the local operator spectrum and central charge of unitary CFTs with a $G=\mathbb Z_N^3$ internal symmetry and LSM anomaly are constrained when there is an additional assumption that the theory possesses certain light defect operators hosted on TDLs of the theory. To obtain central charge bounds, we additionally assume a non-generic gap, the value of which we scan over, to the scaling dimension of any local operator appearing in the OPEs of the defect operators. \par 
To do correlator bootstrap, one must first derive the crossing symmetry constraints arising from all possible four-point functions involving the external operators under consideration. Deriving these constraints by hand is rather cumbersome. Since we know the bootstrap equations for the defect operators of interest are equivalent to those of local operators transforming in $N$-dimensional irreps of $\text{He}(3,N)$, we use the \texttt{autoboot} package by Go and Tachikawa to generate the semidefinite constraints\cite{Go:2019lke}. \texttt{autoboot} is a tool specially designed for producing the crossing symmetry constraints for fields transforming in arbitrary global symmetry representations\footnote{We thank Mocho Go and Yuji Tachikawa for their time in assisting us in setting up \texttt{autoboot} and finding a bug that prevented our correlator bootstrap calculations from working.}. Here we simply summarize the constraints we impose in an abstract form and list the spectrum assumptions used for semidefinite programming; the constraints themselves are generally too unwieldy to report here, but can be made available upon request or simply obtained via \texttt{autoboot}. \par 
The output of \texttt{autoboot} is a set of vector-matrix-valued functions of the conformal cross-ratios $\bold V^{\Delta_D}_{R,\Delta,s}(x,\bar x)$ (i.e. vectors whose components are matrices, with the matrix entries being functions of the conformal cross-ratios), each of which represents the crossing symmetry constraint due to external fields with scaling dimension $\Delta_D$ transforming in $N$-dimensional irreps of $\text{He}(3,N)$ and internal fields transforming in a representation $R$ with scaling dimension $\Delta$ and spin $s$. We take the functions $\bold V^{\Delta_D}_{R,\Delta,s}(x,\bar x)$ to have components expressed in terms of the global conformal blocks for (1+1)$d$ CFTs, given by\cite{Osborn:2012vt,Chester:2019wfx}
\begin{align}
    g_{\Delta,s}(x,\bar x) = \frac{(-2)^{-s}}{1+\delta_{s,0}} (k_{\Delta + s}(x) k_{\Delta - s}(\bar x) + k_{\Delta- s}(x) k_{\Delta+s}(\bar x))
\end{align}
where 
\begin{align}
    k_\beta(x) = x^{\beta/2}{}_2 F_1\left(\frac{\beta}{2}, \frac{\beta}{2}, \beta, x\right).
\end{align}
Each component of the vectors $\bold V^{\Delta_D}_{R,\Delta,s}(x,\bar x)$ corresponds roughly to a distinct four-point function, although the actual implementation in \texttt{autoboot} combines different four-point functions using certain invariant tensors of the symmetry group. If we denote all distinct OPE coefficients, where the internal field is $\mathcal O_R$ and where we use $I,J$ to label different choices of external fields, by $\lambda_{I\mathcal O_R}$ (i.e. $I$ labels a pair of fields here), we can compactly express the crossing symmetry constraints as 
\begin{align}
    \sum_{I,J}\sum_R\sum_{\mathcal O_{R}} \lambda_{I \mathcal O_R}\lambda_{J \mathcal O_R}  [\bold V^{\Delta_D}_{R,\Delta_{\mathcal O_R}, s_{\mathcal O_R}}]^{kIJ} = 0
    \label{defect_crossing_symmetry}
\end{align}
for each $k$, indexing the different individual crossing symmetry constraints. Note that the OPE coefficients in (\ref{defect_crossing_symmetry}) can be chosen to be real\cite{Go:2019lke}.\par 
We now proceed with the usual bootstrap algorithm: we make some assumptions about the spectrum of operators of a given theory and see if our assumptions leads to a violation of (\ref{defect_crossing_symmetry}) using semidefinite programming. \par 
To do the semidefinite programming, we will use \texttt{SDPB} \cite{Simmons-Duffin:2015qma} to search for a (matrix-valued) linear functional acting on the space of vector-matrix-valued functions of $x,\bar x$ that obeys a number of semi-definiteness properties.  For our purposes $\alpha$ may be expressed in a basis of derivatives of the cross ratios evaluated at the crossing-symmetric point $x,\bar x = \tfrac{1}{2}$
\begin{align}
    \alpha [\bold V(x,\bar x)] = \sum_k\sum_{\substack{m,n=0\\m+n \le \Lambda^{\text{cor}}}}^{\Lambda^{\text{cor}}} a^k_{mn} \partial_x^m \partial_{\bar x}^n [\bold V(x,\bar x)]^k \Bigg |_{x=\bar x = \tfrac{1}{2}} \label{correlator_functional}
\end{align}
where $\Lambda^{\text{cor}}$ is the correlator bootstrap derivative order. The properties we need $\alpha$ to obey will depend on whether we are obtaining upper bounds on scaling dimensions or lower bounds on central charge. In the remaining part of this section we explain the constraints on the linear functional in each of these contexts. 
\subsubsection{Upper bounds on scaling dimension}
\begin{figure}[t]
\includegraphics[width=0.5\textwidth]{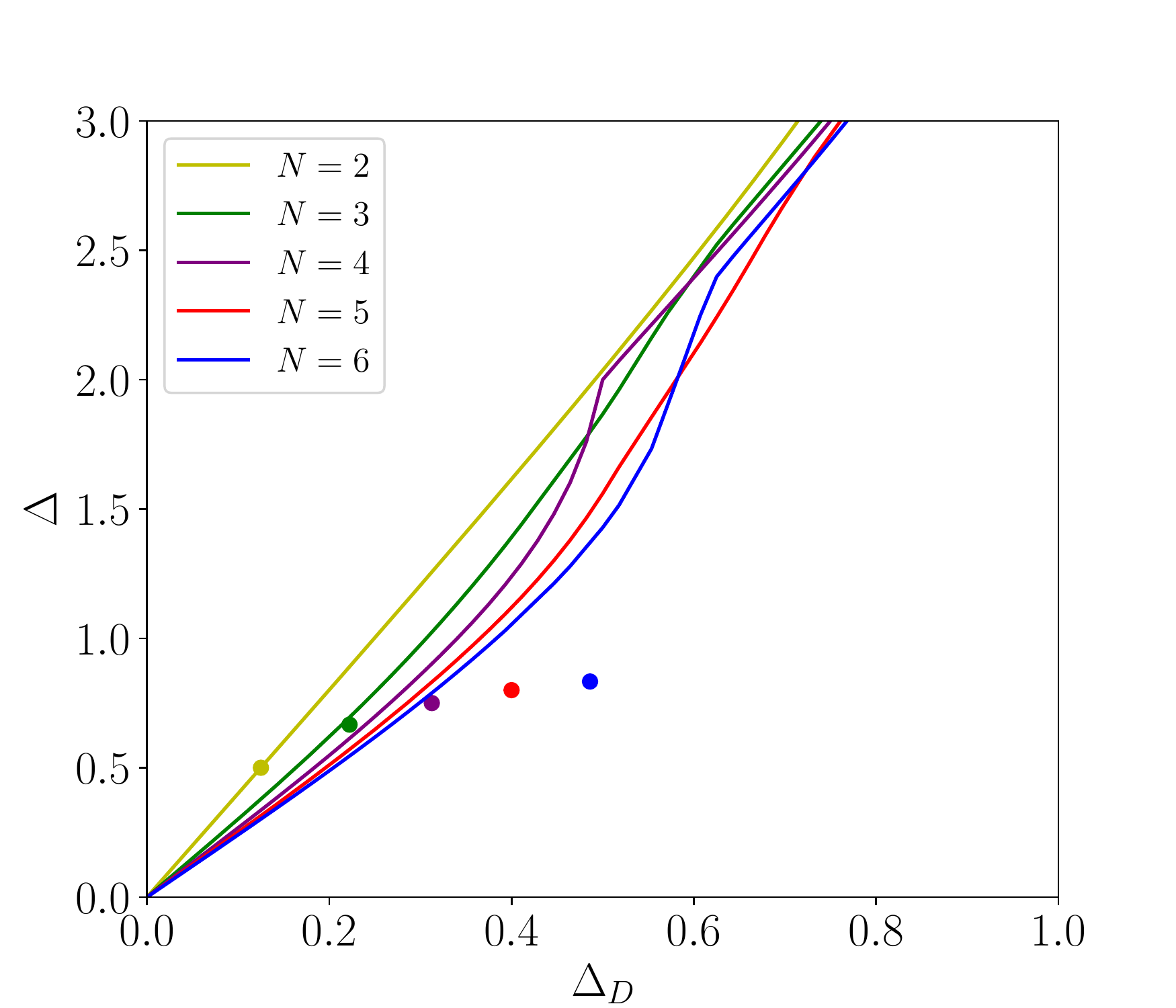}
\caption{Upper bounds on the scaling dimension of the lightest $\mathbb Z_N^3$ charged, scalar operator appearing in the OPE of scalar defect operators with scaling dimension $\Delta_D$ living on a non-anomalous $\mathbb Z_N$ TDL in theories with the LSM anomaly for $N=2,...,6$. The bound is not noticeably changed if the bound is calculated instead for the lightest local operator transforming in \emph{any} representation of $\mathbb Z_N^3$. For each $N$ we assign different colors. The matching colored dots indicate, for each $N$, points corresponding to the scaling dimension of the lightest $\mathbb Z_N^3$-charged local operator appearing in the OPE of the lightest scalar $\mathbb Z_N$ defect operator living on a non-anomalous $\mathbb Z_N$ TDL for $\mathfrak{su}(N)_1$. The $\mathfrak{su}(N)_1$ points for $N=2,...,6$ are $(\tfrac{1}{8},\tfrac{1}{2}),(\tfrac{2}{9},\tfrac{2}{3}),(\tfrac{5}{16},\tfrac{3}{4}),(\tfrac{2}{5},\tfrac{4}{5}),$ and $(\tfrac{35}{72},\tfrac{5}{6})$. These bounds were computed with $\Lambda^{\text{cor}} = 25$ and $S_{\max}^{\text{cor}} = 50$.}
\label{fig:scaling_bounds}
\end{figure}
In order to obtain upper bounds on the scaling dimension of local operators, we will assume that there exist some scalar defect operators $\phi^g_I,\bar \phi^{\bar g}_{\bar I}$ which have the lowest scaling dimension, equal to $\Delta_D$, among all defect operators in the defect Hilbert spaces corresponding to non-anomalous, order $N$ TDLs. Recall that we label irreps of $\text{He}(3,N)$ as $R_i$ according to their index in \texttt{GAP}. For the purposes of bootstrap, we may take these defect operators as transforming in the representation $R_{N^2+1}, \bar R_{N^2+1}$ of $\text{He}(3,N)$ when $N$ is prime or, when $N=4,6$, transforming in $R_{21},R_{54}$ respectively––all such irreps are $N$ dimensional. For brevity later, we will denote the set of all $N$-dimensional irreps by $[N]$. Taking OPEs of these defect operators may produce either local operators or defect operators. This can be seen in the gauged language by observing first that 
\begin{align*}
    R_{N^2 +1} \otimes \bar{R}_{N^2+1} &= \bigoplus_{i = 1}^{N^2} R_i \\
\end{align*}
where on the right-hand side all direct summands are one-dimensional irreps. Note that this is the only possibility for $N=2$ since in this case $R_5$ is a real representation. The one-dimensional irreps can be viewed as representations of $\mathbb Z_N^3$, and thus we see that the anomaly forces the OPE of the defect operators to contain $\mathbb Z_N^3$-charged operators.  For $N > 2$ we also have
\begin{align*}
    &N\text{ prime:} && R_{N^2 + 1} \otimes R_{N^2 + 1} = N R_{N^2+2} \\
    &N = 4: &&  R_{21} \otimes R_{21} = \bigoplus_{i = 17}^{20} 2R_{i} \\
    &N = 6: &&  R_{54} \otimes R_{54} = 3 R_{47} \oplus 3 R_{49} \oplus 3 R_{51} \oplus 3 R_{53}.
\end{align*}
When $N$ is prime, the above tensor products decompose into a direct sum of $N$ copies of another $N$ dimensional irrep, and when $N=4,6$ the decomposition is achieved by a combination of irreps of dimension $N/2$. These tensor product decompositions reflect, in the TDL language, that, for prime $N$, taking the OPE of identical defect operators living on a non-anomalous, order $N$ TDL produces defect operators which also live on order $N$, non-anomalous TDLs. Thus, the scaling dimensions of such operators, when they are scalar, are bounded from below by $\Delta_D$. For our cases $N=4,6$ this is not the case; instead two such defect operators create defect operators living on an order $N/2$ non-anomalous TDL, and we make no assumption about the spectrum of such operators beyond what is guaranteed by unitarity. \par 
We now describe how to rule out various assumptions on the scaling dimension of the lightest operators appearing in the OPE of defect operators with the above listed properties. 
For all $N$ we refer to an operator transforming in \emph{any} non-trivial one-dimensional representation of $\text{He}(3,N)$, which can be interpreted as an operator transforming in a non-trivial representation of $\mathbb Z_N^3$, as a charged operator, and refer to an operator transforming in the trivial representation as a symmetric operator. We will denote the assumed minimum scaling dimensions of any scalar symmetric/charged local operator appearing in the OPE of the defect operators by $\Delta_0^{\text{min}}, \Delta_Q^{\text{min}}$ respectively. We will use $[0]$ and $[Q]$ to denote, respectively, the set of trivial and non-trivial one-dimensional irreps. We will refer to these minimum scaling dimensions $\Delta_0^{\text{min}},\Delta_Q^{\text{min}}$ as the gaps. To attempt to rule out these gaps, we seek a linear functional $\alpha$ of the form (\ref{correlator_functional}), satisfying
\begin{align*}
    \alpha[\bold V_{R_1,0,0}^{\Delta_D}] &= 1 \\
    \alpha[\bold V_{R_i,\Delta,s}^{\Delta_D}] &\succeq 0 &  \forall \Delta \ge \begin{cases} 
    \Delta_0^{\text{min}} & R_i \in [0], s=0 \\
    \Delta^{\text{min}}_{Q} & R_i \in [Q], s=0\\
    \Delta_D & R_i \in [N], s=0, N \text{ odd} \\
    |s| &\text{else} \\
    \end{cases}\\
\end{align*}
where $M \succeq 0$ for a real, symmetric matrix $M$ means that $M$ is positive semidefinite. It is necessary to truncate the values of spin for which we impose the above constraints to only include $|s| \le S^{\text{cor}}_\text{max}$, which we choose to be sufficiently large so that our bounds at fixed $\Lambda^{\text{cor}}$ are stable to an increase in $S^{\text{cor}}_\text{max}$.  Upon finding such a linear functional, we would conclude that the given assumptions are inconsistent with crossing symmetry of the defect operators. If $\Delta_0^{\text{min}} = 0$, we would conclude that $\Delta_Q^{\min}$ is an upper bound on the lightest charged operator appearing in the OPE of the defect operators. If $\Delta_0^{\text{min}} = \Delta_Q^{\text{min}} = \Delta^{\text{min}}$, we conclude that $\Delta^{\text{min}}$ is an upper bound on the lightest operator of any charge. 
\par 
Finally, we will denote by $\Delta^*_Q(N,\Delta_D)$ the optimal, to within some small numerical tolerance, upper bound on the scaling dimension of the lightest charged, local, scalar, operator which appears in the OPE of scalar defect operators transforming in the $R_{N^2+1},\bar R_{N^2+1}$ representations of $\text{He}(3,N)$ with scaling dimension $\Delta_D$. These bounds are shown in Figure \ref{fig:scaling_bounds}. We note that we found the bounds to remain unchanged upon doing similar calculations with $\Delta^{\text{min}}_0 = \Delta^{\text{min}}_Q$, i.e. the resulting bound is on the lightest local operator. We do not present bounds on the lightest symmetric operator since our bounds could not guarantee that the OPE of defect operators must produce a relevant symmetric scalar for any choice of $\Delta_D$, so we do not consider such bounds to be particularly interesting. \par 

\begin{figure*}
    \centering
    \includegraphics[width = 0.49\textwidth]{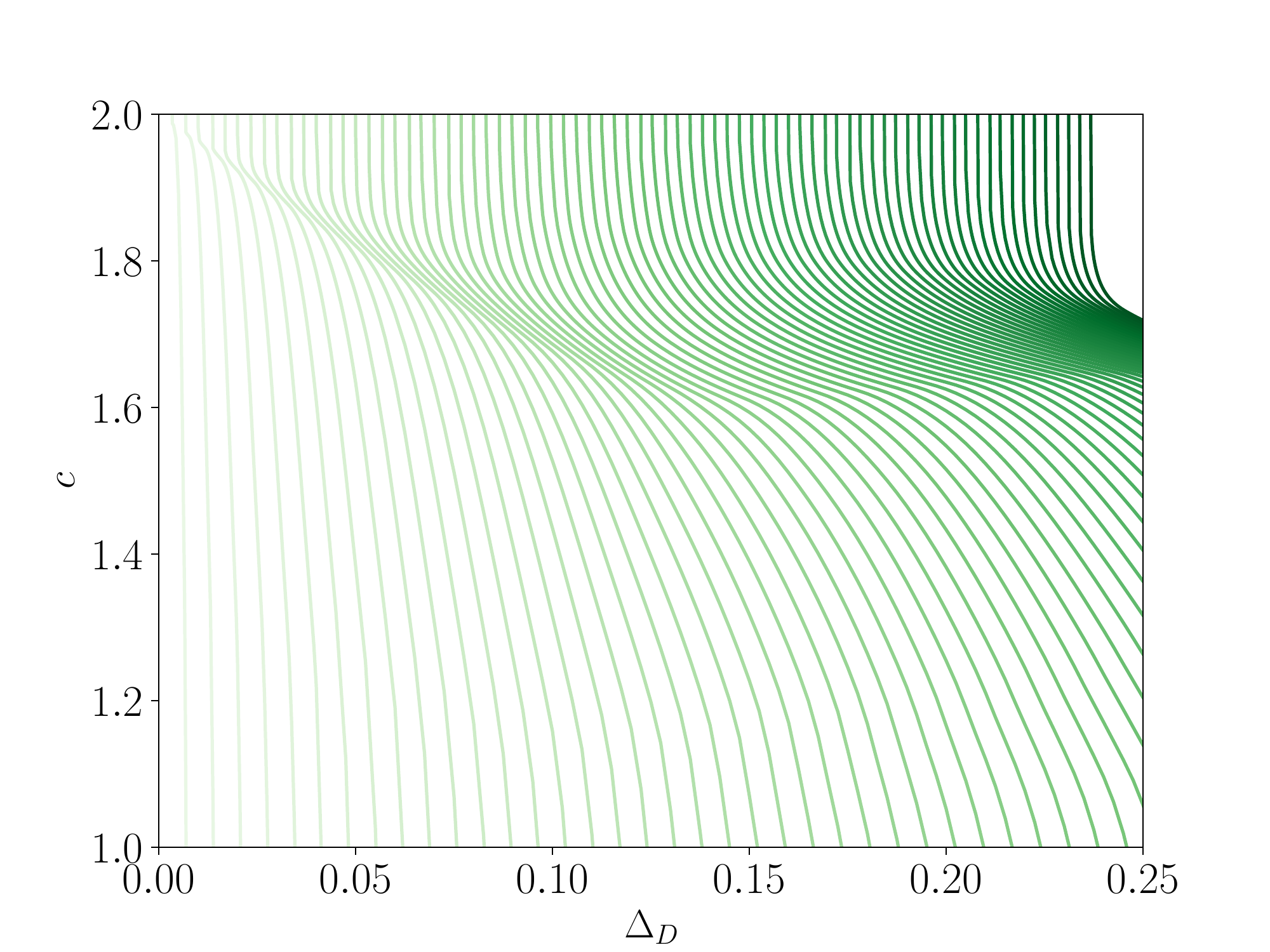}
    \includegraphics[width = 0.49\textwidth]{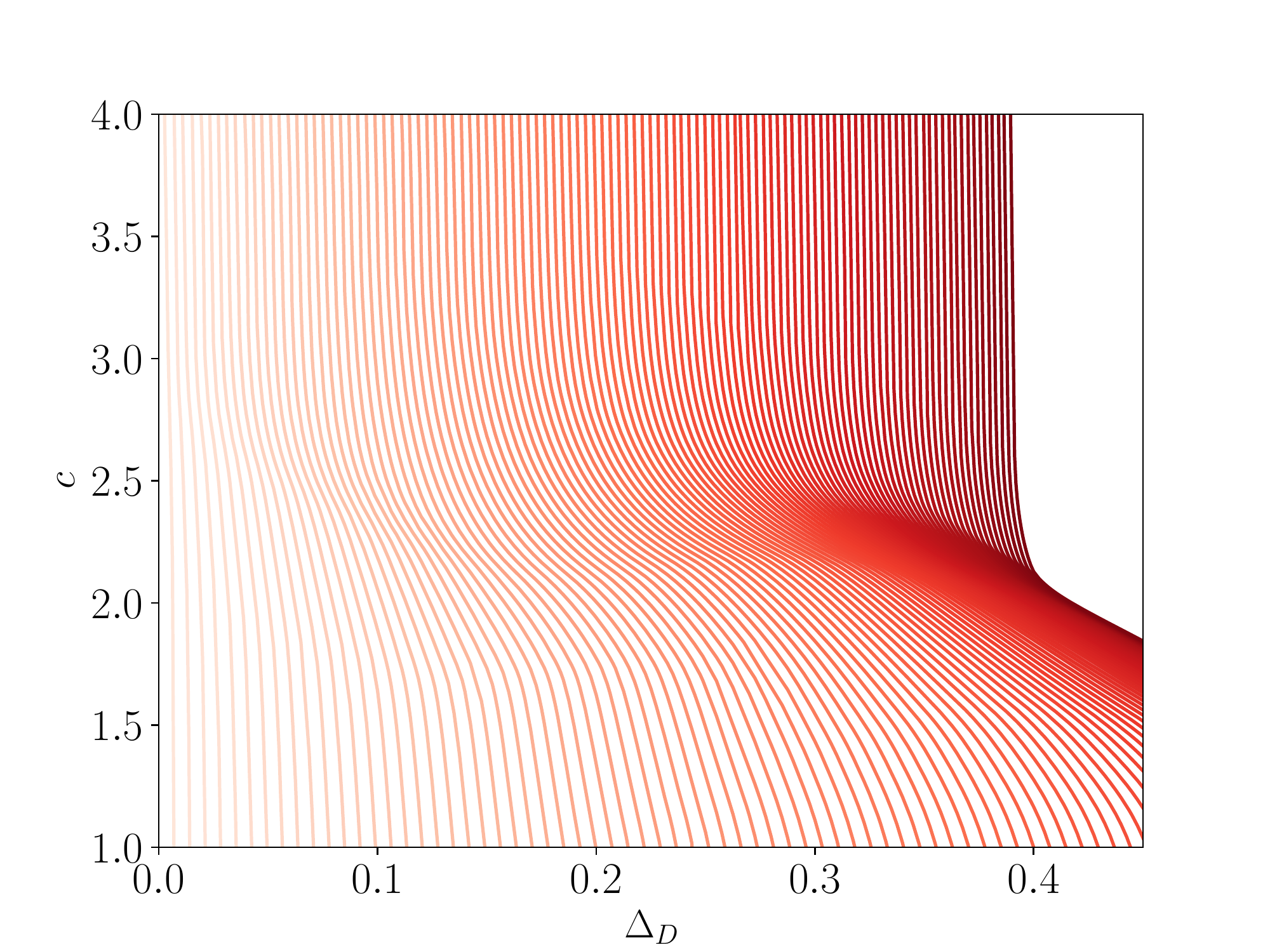}
    \caption{Lower bounds on central charge for $N=3$ (left) and $N=5$ (right) as a function of the scaling dimension of the lightest scalar defect operator living on an order $N$ non-anomalous TDL in a theory with the LSM anomaly. Each solid curve represents a different value of the minimum scaling dimension for the lightest scalar, local operator appearing in the OPE of the external defect operators. For $N=3$ we compute the curves for values where the gap to the minimum value of the scaling dimension for such operators is $\Delta = 0.01,...,0.25$ and, for $N=5$, $\Delta = 0.01,...,1.05$, each with spacing $0.01$ between successive values. The darkness of the curve indicates the value of the gap from low to high with increasing darkness. For $N=3$, the computed values of the central charge lower bound are less than $c=2$, and for $N=5$ all computed values are less than $c=4$, in agreement with known theories. To obtain these bounds we used $\Lambda^{\text{cor}} = 15$ with $S_{\text{max}}^{\text{cor}} = 30$.}
    \label{fig:cent_bounds}
\end{figure*}
\subsubsection{Lower bounds on central charge}
For the central charge bounds, we will assume $N$ is odd and prime since these are the cases for which we are most interested in refining our bounds. \par 
Every unitary, (1+1)$d$ CFT possesses conserved, spin-2, quasiprimary fields which are the holomoprhic and anti-holomorphic stress tensor fields $T(z),\bar T(\bar z)$. With the appropriate normalization\cite{Poland:2018epd}, the OPE coefficient of a primary field $\phi$ with scaling dimension $\Delta$, its conjugate, and the stress tensor is given in 1+1$d$ by\footnote{Note that the extra factor of $\sqrt{\dim R}$ arises due to the definition in \texttt{autoboot} of the OPE coefficient $\lambda_{\phi \bar \phi I} = \sqrt{\dim R}$ for an operator $\phi$ transforming in the representation $R$ of the internal symmetry.}
\begin{align*}
    \lambda_{\phi \bar \phi T} = \frac{2\sqrt{\dim R} \Delta}{\sqrt c},
\end{align*}
where $R$ is the representation of the internal symmetry group that $\phi$ transforms in. In the following we will take $\dim R = N$, applicable to the cases that we consider. \par 
To bound the central charge, we again consider some defect operators $\phi^g_I, \bar \phi^{\bar g}_{\bar I}$ whose scaling dimension is $\Delta_D$. We then expand (\ref{defect_crossing_symmetry}) as
\begin{equation}
\begin{aligned}
   0 = [\bold V_{R_1,0,0}^{\Delta_D}]^k + \lambda^2_{\phi^g \bar \phi^{\bar g} T}[\bold V^{\Delta_D}_{R_1,2,2}]^k \\ + \sum_{I,J} \sum_R \sum_{\substack{\mathcal O_R \ne T,\bar T \\ \Delta_{\mathcal O_R}>0}}\lambda_{I \mathcal O_R} \lambda_{J \mathcal O_R}[\bold V_{R,\Delta_{\mathcal O_R},s_{\mathcal O_R}}^{\Delta_D}]^{kIJ}\big. \label{cbound_crossing_equation}
\end{aligned}
\end{equation}
We now act on (\ref{cbound_crossing_equation}) with a linear functional of the form (\ref{correlator_functional})
\begin{align*}
    \alpha[\bold V_{R_1,2,2}^{\Delta_D}] &= 1 \\
    \alpha[\bold V_{R_i,\Delta,s}^{\Delta_D}] &\succeq 0 &  \forall \Delta \ge \begin{cases}
    \Delta^{\min} \text{ if } R_i \in [0] \cup [Q], s=0\\
    \Delta_D \text{ if  $R_i \in [N]$, $s=0$} \\
    |s| \text{ otherwise} \\
    \end{cases}.\\
\end{align*}
Note that here we have assumed that the scaling dimension of any scalar, local operator appearing in the OPE of the external defect operators has scaling dimension at least $\Delta^{\min}$. It is necessary to impose some gap in the spectrum of symmetric, scalar operators in order to obtain any central charge bounds, since otherwise we would not be able to have $\alpha[\bold V^{\Delta_D}_{R_1,0,0}] < 0$. Acting with such a functional and rearranging terms allows us to conclude 
\begin{align*}
    \lambda_{\phi^g \bar \phi^{\bar g} T}^2 = \frac{4 N \Delta_D^2}{c} \le - \alpha[\bold V_{R_1,0,0}^{\Delta_D}],
\end{align*}
which means, assuming $-\alpha[\bold V_{R_1,0,0}^{\Delta_D}] > 0$ consistent with a gap $\Delta^{\min}$ not being ruled out, 
\begin{align}
    c \ge  \frac{4 N \Delta_D^2}{- \alpha[ \bold V_{R_1,0,0}^{\Delta_D}]}.
\end{align}
Thus, we want to minimize $-\alpha[\bold V_{R_1,0,0}^{\Delta_D}]$ consistent with the previously stated constraints to find the strongest lower bound on $c$, which is again done using \texttt{SDPB}. We will denote our lower bounds obtained in this way, which are presented in Figure \ref{fig:cent_bounds}, by $c^*(N,\Delta_D,\Delta^{\min})$. Central charge bounds in 1+1$d$ CFT have been obtained in other works, e.g. in Ref. [\onlinecite{Vichi:2011zza}], so in order to verify the correctness of our setup, we reproduced some of the central charge bounds contained therein.
\subsection{Modular bootstrap}
\subsubsection{Twisted partition functions}
\begin{figure}[h]
    \centering
    \includegraphics[width=0.15\textwidth]{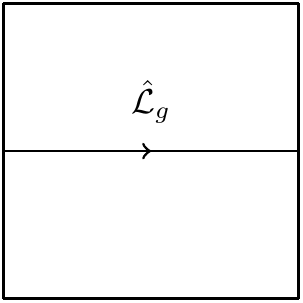}
    \qquad
    \includegraphics[width=0.15\textwidth]{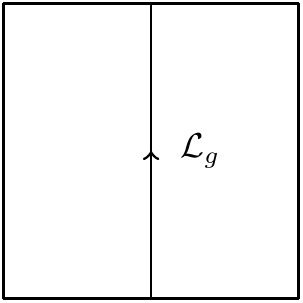}
    \caption{Each square represents a torus, since opposite edges are identified, where we quantize the theory on horizontal spatial slices. On left is the torus partition with a horizontal twist, corresponding to acting with symmetry accross one period of imaginary time evolution. On right is the torus partition function with a vertical twist, corresponding to the partition function of the theory subject to a $\mathcal L_{g}$-twisted boundary condition. }
    \label{fig:twisted_part_funcs}
\end{figure}
The partition function of a unitary, compact, (1+1)$d$ CFT on a spacetime torus encodes the spectrum of scaling dimensions and spins of each primary field of the theory.  When the theory has a global, internal symmetry $G$, we can assign additional quantum numbers accounting for the particular representations $\rho$ of $G$ that states in the Hilbert space of the theory transform in. The torus partition function (which will be denoted pictorially by an empty square with opposite edges identified, i.e. either diagram in Figure \ref{fig:twisted_part_funcs} with the trivial TDL in place of $\mathcal L_g$) with modular parameters $\tau, \bar \tau$ takes the form (when $c > 1$)
\begin{align}
    Z(\tau, \bar \tau) &= \Tr(q^{L_0 - \frac{c}{24}} \bar q^{\bar L_0 - \frac{c}{24}}) \\
    &= \sum_{h,\bar h}\sum_{\rho} n^\rho_{h,\bar h} \, \chi_{h,\bar h}(\tau, \bar \tau)
    \label{torus_part}
\end{align}
where $q = e^{2\pi i \tau},\bar q = e^{-2\pi i \bar \tau}$, $\chi_{h,\bar h}(\tau, \bar \tau) = \chi_h(\tau) \bar \chi_{\bar h}(\bar \tau)$, and $\chi_h$ are the Virasoro characters
\begin{align*}
    \chi_0(\tau) = \frac{(1-q)}{\eta(\tau)}q^{-\frac{c-1}{24}} \qquad \chi_h(\tau) = \frac{1}{\eta(\tau)} q^{h-\frac{c-1}{24}}
\end{align*}
where $\eta(\tau)$ is the Dedekind eta function. Note that to do modular bootstrap calculations, we use so-called reduced characters, defined by replacing $\chi_h(\tau) \to \tau^{1/4} \eta(q) \chi_h(\tau)$. Henceforth, when we write $\chi_h(\tau)$ etc. we will always be referring to these reduced Virasoro characters. In (\ref{torus_part}), $n^\rho_{h,\bar h}$ are positive integers equal to the dimension of the irreducible representation $\rho$ times its multiplicity within each Verma module for conformal dimensions $h,\bar h$.  The assumption of compactness ensures that there is a unique ground state and a discrete spectrum when the spatial extent of spacetime is finite. The partition function is constrained to be invariant under $SL(2,\mathbb Z)$ modular transformations. Such transformations relate descriptions of the same tori with different values of the modular parameters and are generated by the modular $S$ and $T$ transformations 
\begin{align*}
    S : \tau \to -1/\tau \qquad T: \tau \to \tau+1.
\end{align*} 
Imposing equivalence of the torus partition function under such transformations leads to the modular invariance constraints
\begin{align*}
    Z(\tau,\bar \tau) = Z(-1/\tau, -1/\bar \tau) = Z(\tau+1, \bar \tau +1).
\end{align*}
\par 
We can also use TDLs to construct other partition function-like objects that probe aspects of the $G$ global symmetry, which we will call twisted partition functions. These are constructed by considering the theory on a spacetime torus in the presence of TDLs wrapped around the cycles of the torus. When a TDL corresponding to the group element $g$ is wrapped around a spatial cycle, illustrated in the left diagram within Figure \ref{fig:twisted_part_funcs}, this corresponds to acting with $\hat{\mathcal L}_g$ across one period of imaginary time evolution. Specializing to $G= \mathbb Z_N^3$, let us denote the group elements of $\mathbb Z_N^3$ by three-component $\mathbb Z_N$-valued vectors i.e. $g = (i,j,k)$ for some $i,j,k \in \mathbb Z_N$. Similarly, let us denote representations of $G$, which are all one dimensional, in a similar way but, for differentiation, with square brackets $\rho = [i,j,k]$. Then, a state transforming in a representation $\rho$ of $G$ transforms under $g$ as 
\begin{align*}
    \hat{\mathcal L}_g\ket{\rho} = e^{2\pi i \langle \rho, g\rangle/N} \ket{\rho}
\end{align*}
where $\langle \rho,g\rangle = ii'+jj'+kk'$ for $g=(i,j,k)$, $\rho=[i',j',k']$. With this, we may express the torus paritition function with a horizontal twist as 
\begin{align*}
    Z^g(\tau,\bar \tau) &= \Tr_{\mathcal H}\left(\hat{\mathcal L}_g q^{L_0 - \frac{c}{24}} \bar q^{\bar L_0.- \frac{c}{24}}\right) 
    \\ &= \sum_{h,\bar h} \sum_{\rho \in \text{Rep}(G)} n_{h,\bar h}^\rho e^{2\pi i \langle \rho,g\rangle/N}\chi_h (\tau) \bar \chi_{\bar h}(\bar \tau).
\end{align*} 
Likewise, if a $g$-TDL is wrapped around a cycle in the time direction, illustrated within Figure \ref{fig:twisted_part_funcs} on the right, then the resulting twisted partition function represents the partition function of the theory where the trace is taken over the defect Hilbert space $\mathcal H_g$
\begin{align*}
    Z_g(\tau, \bar \tau) &= \Tr_{\mathcal H_g}(q^{L_0 - \frac{c}{24}} \bar q^{\bar L_0 - \frac{c}{24}}) \\ 
    &= \sum_{h,\bar h} n^{g}_{h,\bar h} \chi_{h,\bar h}(\tau,\bar \tau).
\end{align*}
The coefficients of the above twisted partition function $n^g_{h,\bar h}$ are again positive integers representing the degeneracy of each defect operator with conformal dimensions $h,\bar h$. Of course, as we explained previously, the states in the defect Hilbert space may be dressed with (fractionalized) global symmetry quantum numbers as well, but this fact will not be important for our modular bootstrap calculations. \par 
The twisted partition functions are also subject to modular transformation relations. Covariance under $T$ leads to the spin selection rule (\ref{spin-select})\cite{Lin:2019kpn,Lin:2021udi,Chang:2018iay}. Since a modular $S$ transformation swaps the two cycles of the torus it interchanges the space and time directions. Thus, we see that 
\begin{align}
    Z^g(-1/\tau,-1/\bar \tau) &= Z_g(\tau, \bar \tau) \label{mod_S_1}\\
    Z_g(-1/\tau,-1/\bar \tau) &= Z^{\bar g}(\tau,\bar \tau). \label{mod_S_2}
\end{align}
The defect Hilbert space spectrum is thus completely determined by the local operators of the theory and their symmetry representations. Now, we can construct a $2|G|-1$ dimensional vector of twisted partition functions 
\begin{align*}
    \bold Z(\tau,\bar \tau) = \begin{pmatrix}
    Z(\tau,\bar \tau) \\
    Z^g(\tau,\bar \tau) \\
    \vdots \\
    Z_g(\tau,\bar \tau) \\
    \vdots
    \end{pmatrix}.
\end{align*}
This allows us to compactly express the modular transformation relations amongst the twisted partition functions
\begin{align}
    \bold Z(-1/\tau, -1/\bar \tau) =  P \bold Z(\tau,\bar \tau)
\end{align}
where $P$ is a permutation matrix implementing the relations (\ref{mod_S_1},\ref{mod_S_2}) between the components of $\bold Z$.\par 
Borrowing terminology from Refs. [\onlinecite{Lin:2019kpn},\onlinecite{Lin:2021udi}], $\bold Z$ is currently expressed in the \emph{twist basis}, but in order to do bootstrap calculations to obtain bounds which depend on the symmetry representations it is necessary to write $\bold Z$ in the \emph{charge basis}, in which positivity is manifest. This is done by performing a discrete Fourier transformation on the partition functions $Z^g$, leading to partition functions counting the contributions from states transforming in a single representation $\rho$
\begin{align*}
    Z^\rho(\tau, \bar \tau) &= \frac{1}{|G|}\sum_{g \in G} e^{-2\pi i \langle \rho, g \rangle/N} Z^g(\tau, \bar \tau) \\
    &= \sum_{h,\bar h} n^\rho_{h,\bar h} \chi_h(\tau) \bar \chi_{\bar h}(\bar \tau).
\end{align*}
We remind the reader again that the vertically twisted partition functions $Z_g$ already have a decomposition into Virasoro characters with positive coefficients, so no additional change of basis is needed in those instances. Writing $\bold Z(\tau, \bar \tau) =  C \tilde{\bm Z}(\tau, \bar \tau)$, where $\tilde{\bold Z}$ is the twisted partition function in the charge basis and $C$ is a matrix implementing the discrete Fourier transformation, we see that $\tilde{\bm Z}$ transforms under the modular $S$ transformation as
\begin{align*}
    \tilde{\bold Z}(-1/\tau, -1/\bar \tau) = F \tilde{\bold Z} (\tau, \bar \tau) 
\end{align*}
where $F =  C^{-1}  P  C$. \par 
There is an additional step that we perform that greatly simplifies our calculations, which is a generalization of the procedure outlined in Ref. [\onlinecite{Lin:2021udi}] for reducing the dimension of the vector of twisted partition functions $\tilde{\bold Z}$. Given some group $G$, each outer automorphism $\gamma \in \text{Out}(G)$ will generate a reduced vector partition function where the components consist of summing over the twisted partition functions, in the twist basis, corresponding to TDLs which lie in equivalence classes under $\gamma$.  The simplest non-trivial example of this is when $\gamma(g) = g^{-1}$, which pairs each TDL with its orientation reversal, which generates the reduction employed in Ref. [\onlinecite{Lin:2021udi}]. In general, to lose no generality in the bootstrap calculations, one finds the largest subgroup of outer automorphisms, which we will denote $\Gamma \subseteq \text{Out}(G)$, that does not mix TDLs with different $\mathbb Z_N$ anomalies and performs the reduction with the outer automorphisms in this subgroup. We will denote the equivalence classes of TDLs that are related to each other by group automorphisms in $\Gamma$ by $[1], ..., [n_\Gamma]$. That is, denoting elements of $G$ by $g_n$, one finds a reduction matrix 
\begin{align}
    r_{mn} = \begin{cases} 1 & g_n \in [m] \\ 
    0 &\text{else}
\end{cases}.
\end{align}
The full reduction matrix, in the twist basis, is then given by taking two copies of $r_{mn}$, where each copy reduces the horizontally/vertically twisted partition functions (note that the copy for the vertically twisted partition functions excludes the redundant first row of $r$ representing the untwisted partition function)
\begin{align}
    R_{mn} = \begin{cases}
    r_{mn} & 1 \le m \le n_\Gamma, 1 \le n \le |G| \\
    r_{(m-n_\Gamma+1)(n-|G|+1)} & \substack{n_\Gamma+1 \le m \le 2n_\Gamma-1 \\ |G|+1 \le n \le 2|G|-1} \\
    0 & \text{else}
    \end{cases}.
\end{align}
Finally, in the charge basis we write $\tilde R = R C$. Then, defining $\tilde{\bold Z}_{\text{red}} = \tilde R \tilde{\bold Z}$ we may write the simplified modular covariance equation 
\begin{align}
    \tilde{\bold Z}_{\text{red}}(-1/\tau, -1/\bar \tau) = F_{\text{red}}\tilde{\bold Z}_{\text{red}} (\tau,\bar \tau)
\end{align}
where $F_{\text{red}} = \tilde R F \tilde R^T (\tilde R \tilde R^T)^{-1}$. We present $F_{\text{red}}$ for the groups $\mathbb Z_N^3$ we consider in this work in Appendix A. Finally, we define the vectors
\begin{align}
    [\bold M^j_{h, \bar h}(\tau, \bar \tau)]_i =\delta_{ij}\chi_{h, \bar h}(-1/\tau,-1/\bar \tau) - [F_{\text{red}}]_{ij} \chi_{h, \bar h}(\tau, \bar \tau).
\end{align}
Then the statement of modular covariance of the twisted partition function may be expressed as 
\begin{align}
     \sum_{j,h, \bar h} n^j_{h, \bar h}\bold M^j_{h,\bar h}(\tau, \bar \tau) = 0. \label{mod_cov}
\end{align}
\subsubsection{Bounding the local operator spectrum}
To put upper bounds on the scaling dimension of local operators, similarly to correlator bootstrap we aim to show via semidefinite programming that certain assumptions about the spectrum of local and defect operators in a CFT with some global symmetry and anomaly, encoded in the twisted partition functions, lead to a violation of (\ref{mod_cov}). The new ingredients to the standard modular bootstrap program with global symmetries that we introduce are our bounds coming from correlator bootstrap, which allow the more subtle LSM anomaly to enter our final modular bootstrap calculations. We will explain how those bounds lead to non-trivial lower bounds on the scaling dimension of the lightest defect operators, under the assumption of gaps in the spectrum of local operators, and how this can be used to generate universal upper bounds on the scaling dimension of the lightest local operators with various symmetry properties. Incorporating the bounds from correlator bootstrap gives strictly stronger bounds than modular bootstrap alone, and is essential for finding any bound at all on the lightest charged operator when $N$ is odd. \par   
We can make stronger assumptions about the defect operator spectrum due to our previously obtained bounds $\Delta^*(N,\Delta_D)$ and $c^*(N,\Delta_D,\Delta^{\min})$, for reasons we now explain. \par 
As before, let $\Delta_0^{\text{min}}$ and $\Delta_Q^{\text{min}}$ denote the assumed lower bounds on the scaling dimension of any scalar, local operator that is, respectively, symmetric or charged under $\mathbb Z_N^3$. Next define 
\begin{align*}
    \tilde \Delta_D^{\text{scal}}&(\Delta_0^{\text{min}},\Delta_Q^{\text{min}}) \\&\equiv \min\{\Delta_D:\Delta^*(N,\Delta_D) \ge \max(\Delta_0^{\text{min}},\Delta_Q^{\text{min}})\}.
\end{align*}
This represents the minimum scaling dimension of a defect operator living on a non-anomalous, order $N$ TDL whose OPE does not necessarily contain local operators whose scaling dimensions violate the assumed gaps. Thus, $\tilde \Delta_D^{\text{scal}}(\Delta_0^{\text{min}},\Delta_Q^{\text{min}})$ is a lower bound on the scaling dimension of any such defect operator. To incorporate the central charge lower bounds, when $\Delta_0^{\text{min}} = \Delta_Q^{\text{min}} = \Delta^{\text{min}}$, we define 
\begin{align*}
    \tilde \Delta_D^{\text{cent}}(\Delta^{\text{min}},c) \equiv \min \{\Delta_D: c^*(N,\Delta_D,\Delta^{\text{min}}) \le c\}
\end{align*}
which represents the minimum defect operator scaling dimension consistent with gaps $\Delta_0^{\min},\Delta_{Q}^{\text{min}}$ and central charge $c$. Note that, in practice, we can only compute the curves $c^*(N,\Delta_D, \Delta^{\text{min}})$ for a finite list of values $\{\Delta^{(i)}\}$ for the gap in the local operator spectrum $\Delta^{\text{min}}$. Consequently, we perform an interpolation between the central charge curves to achieve the corresponding curve for values of $\Delta^{\text{min}}$ between some $\Delta^{(i)}$ and $\Delta^{(i+1)}$. Given functions $f_i(\Delta_D), f_{i+1}(\Delta_D)$ representing such neighboring curves, we use the interpolation $\tilde f_{i,i+1}(\Delta_D)$ given by
\begin{align*}
    \tilde f_{i,i+1}(\Delta_D) &\equiv f_{i,i+1}^{-1}(\Delta_D) \\
    f_{i,i+1}(c) &\equiv (1-x)f_i^{-1}(c) + x f_{i+1}^{-1}(c) \\
    x &\equiv \frac{\Delta^{\text{min}}-\Delta^{(i)}}{\Delta^{(i+1)} - \Delta^{(i)}}.
\end{align*}
This has the effect of smoothing out our bounds but introduces a small degree of non-rigorousness. Our focus in this work is primarily on showing the existence of a bound and its general, qualitative features, so consequently we do not attempt to quantify the error introduced in this way. However, we should certainly discuss this effect in relation to our claim that $(\mathfrak g_2)_1$ is outside the allowed region in our $N=5$ bound on the lightest local scalar. We can safely rule out $c=\tfrac{14}{5}$ and $\Delta^{\min} = \tfrac{4}{5}$ since the curve $c^*(5,\Delta_D, \tfrac{4}{5})$ is computed exactly, up to discretization effects in $\Delta_D$ that should be very small as the curves are quite smooth. \par
At this point we see that there are two, independent lower bounds on the defect operator scaling dimension that are consistent with either the gaps $\Delta_0^{\text{min}}, \Delta_{Q}^{\text{min}}$ or the central charge $c$; a theory with a defect operator of a lower scaling dimension than the maximum of the two would thus be inconsistent. Finally we define 
\begin{align*}
    \Delta_D^{\text{min}}(\Delta_0^{\text{min}},\Delta_{Q}^{\text{min}},c) \equiv \max\{\tilde \Delta_D^{\text{scal}},  \tilde \Delta_D^{\text{cent}}\}
\end{align*}
and conclude that this is the lowest possible value of the defect operator scaling dimension consistent with the gaps $\Delta_0^{\min}, \Delta_{Q}^{\min}$ and $c$.
\par 
The final step is to do a standard modular bootstrap calculation to determine whether the gaps in the local operator spectrum and the gap in the defect operator spectrum are inconsistent with modular covariance of the twisted partition function. To do this, we seek a linear functional, which acts on vector-valued functions of the modular parameters, of the form
\begin{align}
    \alpha[\bold M(\tau,\bar \tau)] = \sum_j\sum_{\substack{m,n = 0 \\ m+n \le\Lambda^{\text{mod}}}}^{\Lambda^{\text{mod}}} a_{mn}^j \partial_\tau^n \partial_{\bar \tau}^m \left[\bold M(\tau, \bar \tau)\right]_j \Bigg|_{\tau = i, \bar \tau = -i} \label{mod_functional}
\end{align}
where $\Lambda^{\text{mod}}$ is the modular bootstrap derivative order. In our setup, we will allow primary operators of three different types to enter the spectrum of the theories we consider: the vacuum primary with $h=\bar h=0$, degenerate primaries with either $h$ or $\bar h$ being equal to $0$, and non-degenerate primaries with $h, \bar h > 0$. For the case of degenerate primaries and the vacuum, the degenerate Virasoro character of weight 0 enters, while in the non-degenerate case only the non-degenerate characters enter. Further, we will assume that the spectrum of the theories we consider is parity invariant, meaning $n^j_{h, \bar h} = n^j_{\bar h, h}$ without loss of generality, since the anomalies considered in this work are compatible with such an assumption\cite{Anous:2018hjh,Lin:2019kpn,Lin:2021udi}. We thus will define 
\begin{align*}
    \bold M^j_{\Delta,s}(\tau,\bar \tau) \equiv \bold M^j_{h,\bar h}(\tau, \bar \tau) + \bold M^j_{\bar h,h}(\tau, \bar \tau)
\end{align*}
where $\Delta = h+\bar h, s=h-\bar h$ and we assume $s \ge 0$. Note that, in this notation, we can denote the contribution from degenerate Virasoro primaries (including that of the vacuum) by $\bold M_{s,s}^j(\tau, \bar \tau)$ for $s > 0$. The linear functional (\ref{mod_functional}) that we search for will be constrained to have the following properties 
\begin{align*}
    \alpha[\bold M^1_{0,0}] &= 1 \\
    \alpha[\bold M^i_{\Delta,s}] &\ge 0 \qquad \forall \Delta \ge \begin{cases}
    \Delta_0^{\text{min}} & i = 1, s=0 \\
    \Delta_Q^{\text{min}} & 2 \le i \le n_\gamma, s=0 \\ 
    \Delta_D^{\text{min}} & i = 2 n_\gamma -1, s=0 \\
    |s| & \text{else, } s \le S^{\text{mod}}_{\max} \in S_i
    \end{cases} \\
    \alpha[\bold M_{s,s}^i] &\ge 0 \qquad \forall s \le S^{\text{mod}}_{\text{max}} \in S_i
\end{align*}
where we use $S_i$ to denote the spins allowed by the spin selection rule for the local operators, which are just any integer, and defect operators, given in Table \ref{spins_table}, for the $i^{\text{th}}$ composite sector. Note that $2 n_\gamma -1$ is the index denoting the composite defect sector corresponding to non-anomalous, order $N$ TDLs. Further we denote our spin truncation parameter for our modular bootstrap calculations by $S^{\text{mod}}_{\text{max}}$. \par 
Upon finding such a linear functional $\alpha$, we conclude that the assumed gaps $\Delta^{\min}_0$ and $\Delta_Q^{\text{min}}$ are inconsistent with both modular covariance of the twisted partition function and crossing symmetry of defect operators. When we successfully find a linear functional obeying the constraints when $\Delta_0^{\text{min}} = 0$ this represents an upper bound on the lightest charged operator, and similarly when $\Delta_0^{\text{min}} = \Delta_Q^{\text{min}}$ we get a bound on the lightest local operator. Either of these bounds may be optimized, to within a small numerical tolerance, to obtain the final bounds, which are shown in Figure \ref{main_plots}. This concludes our description of our modifications to the usual modular bootstrap procedure. 

\section{Conclusion}
In this work, we take an additional step towards constraining the space of (1+1)$d$ bosonic CFTs with finite global symmetries and anomalies. To this end, we incorporate a more complete picture of the symmetry properties of defect operators into the modular bootstrap, exploiting the delicate balance between the spectrum of defect operators and local operators. We make the relationship between the spectra of defect and local operators quantitatively precise using conformal bootstrap techniques, and in the end obtain universal bounds on the spectrum of local operators. Our primary result is a generalization of the main results of Refs. [\onlinecite{Lin:2019kpn,Lin:2021udi}] that $\mathbb Z_N$ anomalies imply the presence of light, charged operators in a CFT. What we show is that this statement continues to hold for a class of $\mathbb Z_N^3$ anomalies that, in some cases, do not lead to any non-trivial spin selection rule for certain defect operators. The particular symmetries and anomalies that we study occur in physically relevant situations such as spin chains obeying LSM-type constraints, and in somewhat more fine-tuned situations such as multicritical points of (1+1)$d$ SPT phases. Of course, our bounds also apply to the gapless boundary theories of in-cohomology (2+1)$d$ SPT phases protected by $\mathbb Z_N^3$ symmetry with 3-cocycle $\omega$ given by (\ref{lsm_cocycle}). \par 
A question that motivated this work is how, if at all, the central charge of a (1+1)$d$ CFT is bounded from below by its discrete symmetries and anomalies. As far as symmetry goes, it is known that a CFT with a finite global symmetry whose faithfully realized part is a group larger than $\mathbb Z_2$ or $S_3$ must have $c \ge 1$\cite{Ruelle:1998zu}. Further, the Sugawara construction provides a formula for the minimum central charge needed to accommodate continuous global symmetries, which is generally even larger than $c=1$\cite{DiFrancesco}. Thus, restricting to general finite symmetries, we are interested in the question of whether any lower bound $c > 1$ can be proven when the symmetry is anomalous. \par 
Consequently, another goal of our work was to search for numerical bootstrap evidence that could test whether certain suggested lower bounds are reasonable. Some of our calculations seem to suggest that existing, proposed bounds are not quantitatively correct. Specifically, in the case $N=5$ that we study, there is a prominent kink in our plot, but it does not obviously correspond to any known theory. The location of the kink is nearly coincident with the WZW CFT $(\mathfrak g_2)_1$, which has central charge $c=14/5$, but upon doing more careful numerics we were able to place this theory outside of the allowed region. Since bootstrap calculations can only rigorously show what is disallowed, it is difficult to make any statements about features in the allowed region when they do not appear to correspond to known theories. In the future, it will be interesting to explore more carefully the region near this kink to look for other signatures that can explain the kink as an actual theory. We expect that using a mixed correlator bootstrap setup will especially help in this direction, which we are actively investigating. More generally, we are hopeful that future analysis, guided by the evidence provided here, together with additional analytic or numerical calculations, will produce the types of central charge bounds that we desire. \par 
There are various exciting avenues for future study that we would like to mention. First, we can consider technical improvements to the calculations that were done in this work. Among them are, first, using Virasoro conformal blocks for the correlator bootstrap constraints; this would quantitatively improve the bounds and is also a way to directly incorporate the central charge into the correlator bootstrap calculations without needing to assume a non-generic gap in the spectrum of symmetric local operators. Second, we think it would be very interesting to consider the problem of bootstrapping correlation functions of defect operators with fractional spin, which could produce bootstrap bounds on defect operators in the most general setting. The aforementioned improvements would open up the possibility of obtaining current state-of-the-art universal bounds for completely general finite global symmetries and anomalies. This seems especially interesting when the symmetry is described by a fusion category with some non-invertible TDLs\cite{non_inv}. There are countless examples of interesting theories with such non-invertible symmetries, especially in the context of (1+1)$d$ CFT, so exploring the space of CFTs with such symmetries using bootstrap is a natural future direction. \par 
Generalizations of our story to CFTs in higher dimensions seem especially interesting, albeit seemingly in the absence of a useful analog of modular invariance\cite{Belin:2018jtf}. Some progress in this direction has been made already\cite{Lin:2019vgi}, but accounting for the complete picture of symmetries\footnote{Symmetry in QFT in higher dimensions is an active area of study. See i.e. Refs. [\onlinecite{Gaiotto:2014kfa,Cordova:2018cvg,Benini:2018reh,Bhardwaj:2022yxj}] for recent progress.} and anomalies in higher dimensions within bootstrap remains a challenge. There is significant condensed matter theory motivation to incorporate anomalies into bootstrap, since, for instance, it could provide another way to explore the properties of exotic gapless states (see i.e. [\onlinecite{Zou:2021dwv,He:2021sto}]). \par 
How might such higher-dimensional generalizations be achieved? In (1+1)$d$, there is a clear relationship between defect operators and local operators since both sets of operators contribute to the twisted torus partition function, and the OPE of defect operators may contain local operators. In contrast, defect operators corresponding to 0-form global symmetries in dimensions $d$+1 $> 2$ are extended objects, given by conformal defects attached to codimension-1 topological hypersurfaces (for example, in the 3$d$ Ising CFT an example of this is the disorder operator, sometimes known as the twist defect). Such conformal defects host localized operators that have reduced conformal symmetry. Perhaps, then, the implications of anomalies can be reduced to some properties of these operators. There has been much study of such extended objects using a wide range of techniques\cite{PhysRevB.104.L081109,Wang:2021yaf}, including via conformal field theory techniques and bootstrap\cite{Billo:2013jda,Gaiotto:2013nva,Billo:2016cpy}, so including anomalies into the story seems like a natural next step. There are, however, some technical issues that seem to prohibit the most direct generalizations of our work to cases involving extended defects––see i.e. Ref. [\onlinecite{Soderberg:2021kne}]. Additionally, as already pointed out in Ref. [\onlinecite{Lin:2019kpn}], anomalies of finite 0-form symmetries in bosonic CFTs in dimensions $d$+1 $> 2$ can be saturated by gapped topological theories, which have a unique ground state on the spatial sphere. This eliminates the possibility that such anomalies can lead to refined bounds on the spectrum of local operators since any CFT can be made to have any anomaly in this class via stacking gapped degrees of freedom. For such anomalies, it then seems that the questions that bootstrap may be able to answer are those concerning the properties of the extended defects. However, there are several classes of anomalies that ensure gaplessness in higher dimensions, including anomalies for continuous symmetries and even cases involving discrete symmetries\cite{Cordova:2019bsd,PhysRevD.102.025011}. It thus seems promising to continue to develop new ways to incorporate the constraints of such symmetries and anomalies into bootstrap. \par 
\section*{Acknowledgements} 
This work was facilitated through the use of advanced computational, storage, and networking infrastructure provided by the Hyak supercomputer system and funded by the Student Technology Fee at the University of Washington. We thank Tyler Ellison, Justin Kaidi, Ho Tat Lam, Sahand Seifnashri, Shu-Heng Shao, and Nathanan Tantivasadakarn for helpful discussions. We especially thank Tyler Ellison, Justin Kaidi, and Shu-Heng Shao for comments on an initial draft of this work. We thank Mocho Go and Yuji Tachikawa for their assistance with \texttt{autoboot}. RL thanks the 2021 annual meeting for the Simons Collaboration on Non-Perturbative bootstrap, as well as the Simons Center for Geometry and Physics at Stony Brook University, for hospitality. RL and LF are supported by NSF DMR-1939864.

\par 
\appendix 
\section{Modular bootstrap equations}
The reduced modular covariance equation reads 
\begin{align*}
     \tilde{\bold Z}_{\text{red}}(-1/\tau, -1/\bar \tau) = F_{\text{red}}\tilde{\bold Z}_{\text{red}} (\tau,\bar \tau)
\end{align*}
Here we list $F_{\text{ref}}$ for each $N$ along with the corresponding reduced partition function components. 
\subsection{$N \in 2 \mathbb Z + 1$}
As discussed, when $N$ is odd the LSM anomaly does not lead any $\mathbb Z_N$ subgroup to have a $\mathbb Z_N$ anomaly. When $N$ is prime, such as in the cases we consider, the full orbit of the automorphism group of any non-trivial $g \in \mathbb Z_N^3$ contains all other non-trivial TDLs. Consequently, we may perform the maximal reduction and lose no generality. If we denote the trivial representation by $\rho_1$ and the trivial element of $G$ by $g_1$ then 
\begin{align*}
    F_{\text{red}} = \begin{pmatrix}
    \frac{1}{|G|} & \frac{1}{|G|} & \frac{1}{|G|} \\
    \frac{|G|-1}{|G|} & \frac{|G|-1}{|G|} & - \frac{1}{|G|} \\
    |G|-1 & -1 & 0
    \end{pmatrix}
&&
    \tilde{\bold Z}_{\text{red}} = \begin{pmatrix}
    Z^{\rho_1} \\ 
    \sum_{\rho_1 \ne \rho \in \text{Rep}(G)}Z^\rho \\
    \sum_{g_1 \ne g \in G } Z_g 
    \end{pmatrix}
\end{align*}
\subsection{\texorpdfstring{$N \in 2 \mathbb Z$}{N in 2 Z}}
In the case of even $N$, the full reduction is no longer possible. However, by grouping together all TDLs with identical $\mathbb Z_M$ anomalies according to Table \ref{spins_table} we still may perform a non-trivial reduction.
\subsubsection{\texorpdfstring{$N=2$}{N=2}}
According to Table \ref{spins_table} there are two types of TDLs, both with order 2: non-anomalous or anomalous with $k=1$. Denote the subset of $G$ corresponding to the former as $[2\text{N}]$ and the latter as $[2\text{A}]$. \par 
Next, denote representations of $\mathbb Z_N^3$ as vectors $\rho = [i,j,k]$. There are then two automorphism classes of representations induced by automorphisms preserving the spin selection rules of the LSM anomaly: a class $[\text{O}]$ where an odd number of components of $\rho$ is non-trivial and a class $[\text{E}]$ where an even number is non-trivial. Then
\begin{align*}
\renewcommand\arraystretch{1.3}
F_{\text{red}} = \begin{pmatrix} 
\frac{1}{8} & \frac{1}{8} & \frac{1}{8} & \frac{1}{8} & \frac{1}{8} \\
\frac{3}{8} & \frac{3}{8} & \frac{3}{8} & \frac{3}{8} & -\frac{1}{8} \\
\frac{1}{2} & \frac{1}{2} & \frac{1}{2} & - \frac{1}{2} & 0 \\
1 & 1 & -1 & 0 & 0 \\
6 & -2 & 0 & 0 & 0
\end{pmatrix}
&&
    \tilde{\bold Z}_{\text{red}} = \begin{pmatrix}
    Z^{\rho_1} \\ 
    \sum_{\rho \in [\text{E}]}Z^\rho \\
    \sum_{\rho \in [\text{O}]}Z^\rho \\
    \sum_{g \in [2\text A] } Z_g \\
    \sum_{g \in [2\text N] } Z_g
    \end{pmatrix}
\end{align*}
\subsubsection{\texorpdfstring{$N=4$}{N=4}} 
Table \ref{spins_table} indicates that there are three classes of TDLs. All order 2 TDLs are non-anomalous and we will denote the class of such TDLs by $[2\text{N}]$. Order 4 TDLs are either anomalous with $k=2$ or non-anomalous, respectively denoted by $[4\text{A}]$ and $[4\text N]$. There are also three classes of non-trivial representations. The first, denoted $[2 \text E]$, are representation vectors where an even number of components is equal to 2 and the rest are equal to 0. Then $[2 \text O]$ denotes the representation vectors where an odd number of components are equal to 2 and the rest are 0. Finally, $[4]$ denotes all remaining, non-trivial representation vectors. These give 
\begin{widetext}
\begin{align*}
\renewcommand\arraystretch{1.3}
    F_{\text{red}} = \begin{pmatrix}
    \frac{1}{64} & \frac{1}{64} & \frac{1}{64} & \frac{1}{64} & \frac{1}{64} & \frac{1}{64} & \frac{1}{64} \\
    \frac{3}{64} & \frac{3}{64} & \frac{3}{64} & \frac{3}{64} & \frac{3}{64} & \frac{3}{64} & -\frac{1}{64} \\
    \frac{1}{16} & \frac{1}{16} & \frac{1}{16} & \frac{1}{16} & \frac{1}{16} & -\frac{1}{16} & 0 \\
    \frac{7}{8} & \frac{7}{8} & \frac{7}{8} & \frac{7}{8} & - \frac{1}{8} & 0 & 0 \\
    7 & 7 & 7 & -1 & 0 & 0 & 0 \\
    8 & 8 & -8 & 0 & 0 & 0 & 0 \\
    48 & -16 & 0 & 0 & 0 & 0 & 0
    \end{pmatrix}
&&
    \tilde{\bold Z}_{\text{red}} = \begin{pmatrix}
    Z^{\rho_1} \\ 
    \sum_{\rho \in [\text{2E}]}Z^\rho \\
    \sum_{\rho \in [\text{2O}]}Z^\rho \\
    \sum_{\rho \in [\text{4}]}Z^\rho \\
    \sum_{g \in [2\text N] } Z_g \\
    \sum_{g \in [4\text A] } Z_g \\
    \sum_{g \in [4\text N] } Z_g
    \end{pmatrix}
\end{align*}
\end{widetext}
\subsubsection{\texorpdfstring{$N=6$}{N=6}} 
Following a similar pattern as before, we label the classes of TDLs as $[2\text N],[2\text A],[3\text N],[6\text N],[6 \text A]$ where the order 2 anomalous TDLs have $k=1$ and the order 6 anomalous TDLs have $k=3$. There are five classes of non-trivial representations. We denote the classes containing an even/odd number of components being equal to 3 as $[3\text E]/[3 \text O]$. There is a class where each component is an even number, which we denote by $[\text E]$. There is a class containing only odd numbers with at least one component not equal to 3, denoted $[\text O]$. Finally, there is a class where the components are a mix of even and odd numbers, denoted $[\text{EO}]$. These give
\begin{widetext}
\begin{align*}
\renewcommand\arraystretch{1.3}
\setcounter{MaxMatrixCols}{20}
    F_{\text{red}} = \begin{pmatrix}
    \frac{1}{216} & \frac{1}{216} & \frac{1}{216} & \frac{1}{216} & \frac{1}{216} & \frac{1}{216} & \frac{1}{216} & \frac{1}{216} & \frac{1}{216} & \frac{1}{216} & \frac{1}{216} \\
    \frac{1}{72} & \frac{1}{72} & \frac{1}{72} & \frac{1}{72} & \frac{1}{72} & \frac{1}{72} & \frac{1}{72} & -\frac{1}{216} & \frac{1}{72} & \frac{1}{72} & -\frac{1}{216} \\ 
    \frac{1}{54} & \frac{1}{54} & \frac{1}{54} & \frac{1}{54} & \frac{1}{54} & \frac{1}{54} & -\frac{1}{54} & 0 & \frac{1}{54} & - \frac{1}{54} & 0 \\
    \frac{13}{108} & \frac{13}{108} & \frac{13}{108} & \frac{13}{108} & \frac{13}{108} & \frac{13}{108} & \frac{13}{108} & \frac{13}{108} & -\frac{1}{216} & -\frac{1}{216} & -\frac{1}{216} \\
     \frac{13}{36} &  \frac{13}{36} &  \frac{13}{36} &  \frac{13}{36} &  \frac{13}{36} &  \frac{13}{36} &  \frac{13}{36} & - \frac{13}{108} & - \frac{1}{72} & -\frac{1}{72}& \frac{1}{216} \\
     \frac{13}{27} & \frac{13}{27} & \frac{13}{27} & \frac{13}{27} & \frac{13}{27} & \frac{13}{27} & -\frac{13}{27} & 0 & -\frac{1}{54} & -\frac{1}{54} & 0 \\
     1 & 1 & -1 & 1 & 1 & -1 & 0 & 0 & 0 & 0 & 0 \\
     6 & -2 & 0 & 6 & -2 & 0 & 0 & 0 & 0 & 0 & 0 \\
     26 & 26 & 26 & -1 & -1 & -1 & 0 & 0 & 0 & 0 & 0 \\
     26 & 26 & -26 & -1 & -1 & 1 & 0 & 0 & 0 & 0 & 0 \\
     156 & -52 & 0 & -6 & 2 & 0 & 0 & 0 & 0 & 0 & 0 
    \end{pmatrix}
    && \tilde{\bold Z}_{\text{red}} = \begin{pmatrix}
    Z^{\rho_1} \\
    \sum_{\rho \in [3\text E]} Z^\rho \\
    \sum_{\rho \in [3\text O]} Z^\rho \\
    \sum_{\rho \in [\text E]} Z^\rho \\
    \sum_{\rho \in [\text O]} Z^\rho \\ 
    \sum_{\rho \in [\text{EO}]} Z^\rho \\
    \sum_{g \in [2\text A]} Z_g \\
    \sum_{g \in [2 \text N]} Z_g \\
    \sum_{g \in [3 \text N]} Z_g \\
    \sum_{g \in [6\text A]} Z_g \\
    \sum_{g \in [6\text N]} Z_g \\
    \end{pmatrix}
\end{align*}
\end{widetext}

\bibliography{bibliography.bib}

\end{document}